\documentclass[aps,pra,reprint,superscriptaddress,longbibliography]{revtex4-2}

\usepackage{amsmath, amsthm, amsfonts, amssymb,amstext}
\usepackage{mathptmx}
\usepackage{bm}
\usepackage{xfrac}
\usepackage{mathrsfs}
\usepackage{latexsym}
\usepackage{array}

\usepackage{soul}

\usepackage{graphicx}

\usepackage{amsmath,amssymb,amsfonts}

\usepackage{enumitem}

\usepackage{xcolor}

\usepackage{soul,xcolor}

\usepackage{tikz,tkz-euclide}

\usepackage[colorlinks=true,citecolor=blue,linkcolor=blue,urlcolor=blue]{hyperref}

\newcommand{\abs}[1]{\left|#1\right|}

\let\ni\noindent
\usepackage{physics}
\usepackage{verbatim}
\usepackage{array}
\usepackage{tabularx}

\usepackage{tikz,ifthen} 
\usetikzlibrary{shapes,arrows,positioning,automata,backgrounds,calc,er,patterns,matrix}

\begin{document}
\begin{abstract}
With the advent of quantum simulators, exploring exotic collective phenomena in  lattice models with local symmetries and unconventional geometries is at reach of near-term experiments.
Motivated by recent progress in this direction, we study a $\mathbb{Z}_2$ lattice gauge theory defined on a multi-graph with links that can be visualized as great circles of a  spherical shell  hosting the $\mathbb{Z}_2$ gauge fields. Elementary Wilson loops along pairs of these bonds allow to identify a dynamical gauge-invariant flux, responsible for Aharonov-Bohm-like interference effects in the tunneling dynamics of charged matter residing on the vertices. Focusing on an odd number of links, we show that this leads to state-dependent tunneling amplitudes underlying a phenomenon analogous to the Peierls instability. We find  inhomogeneous phases  in which an ordered pattern of the gauge fluxes spontaneously breaks translational invariance, and intertwines with a bond order wave for the gauge-invariant kinetic matter operators. Long-range order is shown to coexist with symmetry protected topological order, which survives the quantum fluctuations of the gauge flux induced by an external electric field. Doping the system above half filling leads to the formation of topological soliton/anti-soliton pairs interpolating between different inhomogeneous orderings of the gauge fluxes. By performining a detailed analysis based on matrix product states, we prove that charge deconfinement emerges as a consequence of charge-fractionalization. Quasiparticles carrying fractional charge and bound at the soliton centers can be arbitrarily separated without feeling a confining force, in spite of the long-range attractive interactions set by the small electric field on the individual integer charges.
\end{abstract}

\title{Symmetry-protected topology and deconfined solitons  in  a multi-link $\mathbb{Z}_2$ gauge theory}

\author{Enrico C. Domanti}

\email{Enrico.Domanti@gmail.com}
\affiliation{Instituto de Física Teorica, UAM-CSIC, Universidad Autonoma de Madrid, Cantoblanco, 28049 Madrid, Spain}

\author {Alejandro Bermudez}

\affiliation{Instituto de Física Teorica, UAM-CSIC, Universidad Autonoma de Madrid, Cantoblanco, 28049 Madrid, Spain}

\maketitle

\setcounter{tocdepth}{2}
\begingroup
\hypersetup{linkcolor=black}
\tableofcontents
\endgroup

\begin{section}{\bf Introduction}

Lattice gauge theories (LGTs) are traditionally formulated by placing  matter fields on the sites and  gauge fields  on the links of an hyper-cubic lattice that discretises spacetime, such that  there is a single gauge-field link per pair of nearest-neighbor matter sites~\cite{PhysRevD.10.2445,kogut1983lattice}. The standard view is that  the lattice is  an artificial scaffolding, and one should carefully remove  the associated  artifacts when approaching the continuum limit~\cite{carrol1976lattice,sharatchandra1978continuum}, the price to pay in order to explore the strongly-coupled properties of LGTs numerically. LGTs have also become the focus of attention in quantum simulators~\cite{Feynman1982}, namely  controlled many-body systems capable of mimicking gauge-field dynamics in a controlled playground~\cite{buchler2005atomic,Weimer2010rydberg,zohar2011confinement,zohar2012simulating,zohar2013cold,zohar2013simulating,Zohar2016quantum,banerjee2012atomic,banerjee2013atomic,wang2025quantum,banuls2020review,Banuls2020simulating,tagliacozzo2013optical,Tagliacozzo2013simulation,Wiese2013ultracold,dalmonte2016lattice,Aidelsburger2021cold,popov2024variational,gaz2025quantum,popov2025nonperturbative}. This focus has led to the observation of relevant phenomena such as confinement and string breaking~\cite{Martinez2016real,Dai2017four,klco2018quantum,Schweizer2019floquet,Kokail2019self,surace2020lattice,klco2020su2,mil2020scalable,Yang2020observation,zhao2022thermalization,Atas2021su2,bauer2021simulating,nguyen2022digital,ciavarella2021trailhead,mildenberger2025probing,rahman2021su2,ciavarella2022preparation,wang2022observation,atas2023simulating,farrell2023preparations,farrell2023preparations2,su2023observation,charles2024simulating,zhang2023observation, Cochran2025visualizing,de2024observationstringbreakingdynamicsquantum,gonzalez2025observation,liu2024stringbreakingmechanismlattice,saner2025real,cobos2025realtimedynamics21dgauge,halimeh2025quantum,gyawali2024observation,schuhmacher2025observation}, which  calls for a new perspective for LGTs: the lattice is a physical reality, and   future advances may allow to design unconventional lattice geometries and link structures at will, allowing one to explore novel effects that go beyond the original LGT scenario.
With this mindset, features once dismissed as discretization artifacts can now become relevant and lead to new phenomena, unveiling a yet uncharted territory  for gauge theories. 

In this paper, we introduce an apparently-simple  gauge theory on a non-standard lattice: a (1+1)-dimensional chain of  fermionic $\mathbb{Z}_2$ charges coupled to a $\mathbb{Z}_2$ gauge field through   gauge-invariant tunnelings along multiple links and, thus, a LGT that is mathematically defined on an undirected multi-graph (see Fig.~\ref{fig:z2_scheme}{\bf (a)}). In this lattice model,  one can identify various gauge-invariant Wilson loops   encoding flux configurations that can take two possible values, either $0$ or $\pi$. These fluxes can be  pictorically understood  as the result of the dynamical $\mathbb{Z}_2$  field piercing the  spherical caps enclosed by the links  (see Fig.~\ref{fig:z2_scheme}{\bf (b)}). This raises a potentially interesting interplay of quantum interference effects for the dynamics of the $\mathbb{Z}_2$ charges, reminiscent of the Aharonov–Bohm phenomenon~\cite{PhysRev.115.485}, with  effects caused by the  quantum fluctuations of the gauge flux controlled by the ratio of  electric- and   magnetic-type terms. The goal of this work is to study the intertwining of  matter and gauge fields, exploring new effects in which spontaneous symmetry breaking (SSB) and symmetry-protected topology (SPT)~\cite{RevModPhys.88.035005} become manifest.  

\begin{figure*}
  \centering
  \includegraphics[width=1.9\columnwidth]{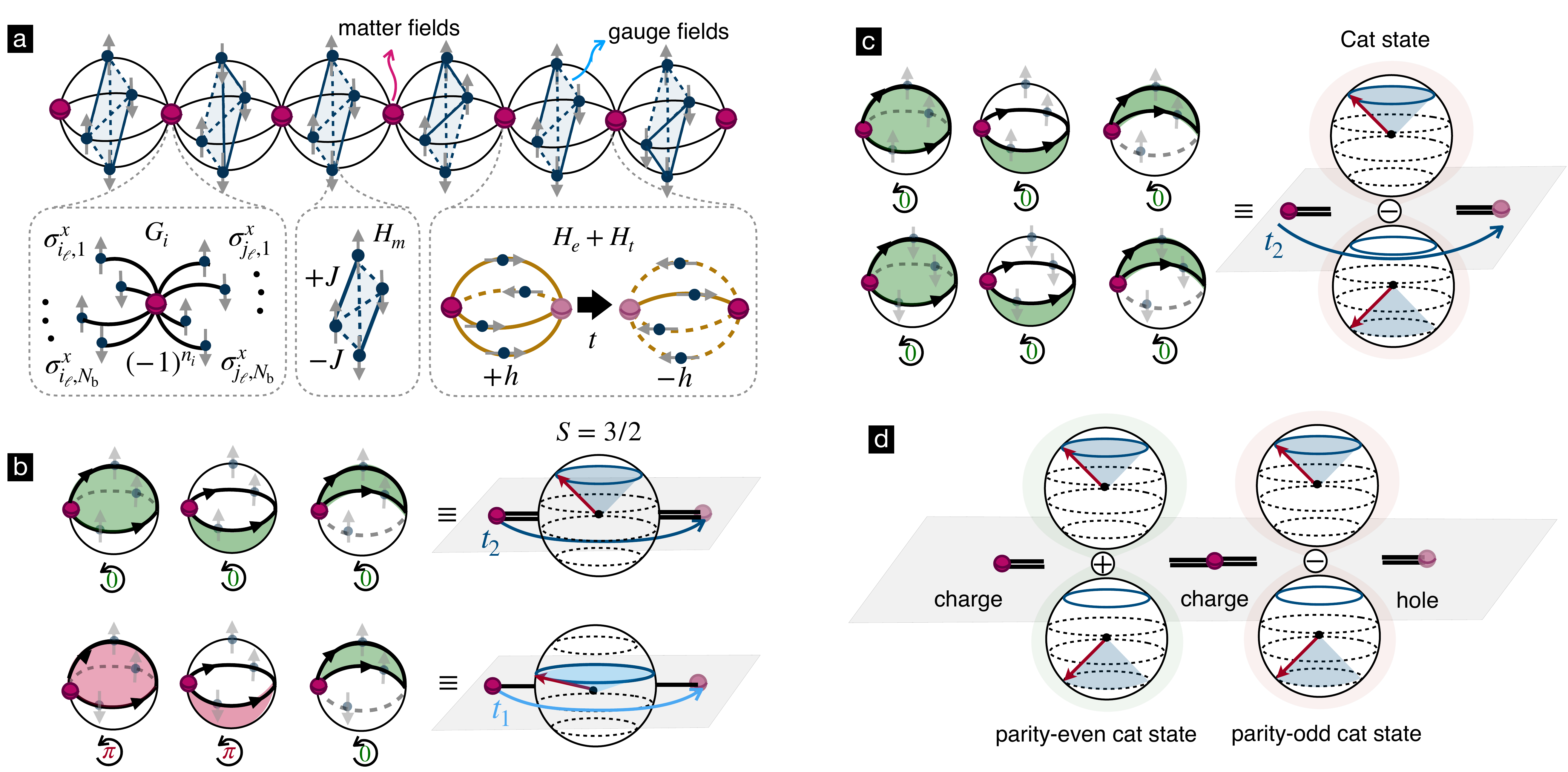}
  \caption{\textit{Multi-link $\mathbb{Z}_2$ gauge theory}. (a) Pictorial representation of the multi-link chain: fermions live on the lattice sites (red dots) while gauge fields are spin-$\frac{1}{2}$ Pauli matrices located on its bonds, which are depicted as halves of great circles on a sphere of diameter equal to the lattice spacing $a$. The $\mathbb{Z}_2$ gauge generators $G_i$ have support on each lattice site and on the neighboring bonds belonging to the adjacent spherical shells - see Eq.~\eqref{eq:generators}. The Hamiltonian~\eqref{eq:model} consists of a magnetic term $H_m$ comprising all possible pairwise antiferromagnetic interactions between spins, controlled by a parameter $J$, of the gauge invariant tunneling $H_t$ and the electric field contribution $H_e$, respectively proportional to the parameters $t$ and $h$. Particles can hop along any of the $N_b$ bonds connecting neighboring sites, followed by a flip of the corresponding spin state (in the Hadamard basis, represented by arrows directed along each bond) to fulfill gauge invariance. (b) In the case $N_b=3$, working in the maximal spin $S=\frac{3}{2}$ sector of the total spin operator $\boldsymbol{S}_{i_\ell}$, the eigenstates $S^z_{i_\ell}$ corresponding to $m=\frac{3}{2},\frac{1}{2}$ are illustrated, on the left, in terms of their spin-$\frac{1}{2}$ components on each bond. Each spherical cap subtended by a pair of links is threaded by a gauge-flux $\langle W^{a,b}_{i_\ell,\circlearrowleft}\rangle = \langle \sigma^z_{i_\ell,a} \sigma^z_{i_\ell,b}\rangle$ taking values $0$ or $\pi$, respectively shown in green or red and corresponding to a ferromagnetic or antiferromagnetic ordering of the spins on the two bonds. On the right, these states are represented by shaded blue circles at a fixed value of the z-projection of $\langle \boldsymbol{S}_{i_\ell}\rangle$, illustrated as a red vector originating from the center of the sphere. While the two different eigenstates of $S^z_{i_\ell}$ would result in different tunneling amplitudes in a naive mean-field treatment of Eq.~\eqref{eq:totspin}, gauge invariance forces $\langle S^z_{i_\ell}\rangle=0$ on physical states, emphasizing the need for a different local spin basis to discuss state-dependent tunneling amplitudes - see Eq.~\eqref{eq:good_basis}. (c) Cat state $\ket{\frac{3}{2},-}=\frac{1}{\sqrt{2}}(\ket{\frac{3}{2}} - \ket{-\frac{3}{2}})$ supporting tunneling with the amplitude $t_2 = \frac{3}{2} t$. (d) Gauss-law constraints on the distribution of $\mathbb{Z}_2$ charges and parity cat states of the gauge fields at the multi-links. } 
  \label{fig:z2_scheme}
\end{figure*}

In a standard LGT context, the (1+1)-dimensional $\mathbb{Z}_2$ LGT is instead defined on an undirected graph with a single gauge link that connects a pair of neighboring matter sites, allowing for a gauge-invariant tunneling term. This leads to a featureless phase diagram: it is generically confining and does not support a topological phase through deconfinement~\cite{borla2020confined,kebric2021confinement}. Long-range order in this model is  precluded by Elitzur's theorem~\cite{elitzur1975impossibility}, as  local gauge symmetries cannot be spontaneously broken and its groundstate is a Luttinger liquid of  charge-neutral dimers confined by  electric-field lines, which play the role of mesons in higher-dimensional LGTs. In the present work, we show that this behavior changes completely when moving to multi-graphs with an odd number of links connecting two sites: the interplay of Aharonov–Bohm interference, quantum fluctuations and gauge dynamics stabilizes translation–symmetry–broken patterns of the gauge field which, in turn, induce topological phases and soliton-like topological excitations in the matter sector. For multi-graphs with $N_b = 3$ links/bonds (see Fig.~\ref{fig:z2_scheme}{\bf (b)}), we demonstrate a direct connection of our LGT to the Peierls mechanism of lattice distortion in 1D metals~\cite{peierls2001quantum}. We demonstrate that a spontaneous breakdown of  translational invariance manifests in the periodic modulation of certain gauge-invariant observables characterized by a wave-vector that depends on the particle filling $\nu$. We show that the ordered phase emerging from {SSB} coexists with an {SPT} phase~\cite{chen2011classification,chen2012symmetry,wen2012symmetry,senthil2015symmetry,chiu2016classification},  particularly associated to  the $\mathsf{BDI}$ symmetry class~\cite{chiu2016classification}. Moreover, when doping the system above specific commensurate fillings, we demonstrate a mechanism of fractionalization-induced deconfinement in which the additional $\mathbb{Z}_2$ charges fractionalize into new quasi-particles localised at topological defects. Such defects are generated by the solitonic configurations of the gauge-field SSB patterns, which are free to move carrying the fractional charges without feeling a linearly-rising confining force, nor eventually leading to string-breaking phenomena.

A motivation to study this type of multi-graph gauge theories follows from recent experimental advances that have managed to reproduce $N_b=2$ links on a single $\mathbb{Z}_2$ loop, showing the interplay of gauge-field dynamics with an Aharonov-Bohm effect in analog quantum simulators based on trapped ions~\cite{saner2025real}. An interesting feature  of these particular multi-link models is that the Wilson loops do not require complicated high-weight plaquette interactions to respect the local gauge symmetry. Instead, the Wilson loops are weight-2 operators, and the corresponding magnetic-type terms of the Hamiltonian are two-body Ising interactions which have been demonstrated in several seminal experiments with trapped ions~\cite{Friedenauer:2008gny,RevModPhys.93.025001}. Accordingly, the LGTs explored in this work lead to a simpler competition of magnetic and electric terms intertwined with the charge dynamics, and might be at reach of near-term analog quantum simulators  before the fault-tolerant digital counterparts become available.

This article is organized as follows. In Sec.~\ref{sec:model} the Hamiltonian of the multi-link $\mathbb{Z}_2$ gauge theory is presented and its local gauge symmetry is discussed. In Sec.~\ref{sec:peierls} we explore the interplay between the particles' dynamics and the configuration of elementary Wilson loops, leading to a Peierls transition and to the SSB of translational invariance, manifesting in an inhomogeneous bond-ordered wave. In Sec.~\ref{sec:tbow}, the latter is characterized as an SPT phase and its extent is delimited through a finite-size scaling analysis based on thorough numerical simulations based on matrix product states (MPS). The emergence of solitonic defects in the profile of gauge fluxes is addressed in Sec.~\ref{sec:solitons}, particularly by doping the system with extra particles above half filling. Remarkably, we show that quasi-particles with fractional charge emerge, which are bound to a topological soliton/anti-soliton configuration. Moreover, these quasi-particles  are deconfined, as they can be pulled apart to arbitrarily-large distances without feeling any confining force.

\end{section} 

\begin{section}{\bf  $\mathbb{Z}_2$ lattice gauge theories on multigraphs}
\label{sec:model}

We consider a  Hamiltonian formulation of the multi-link $\mathbb{Z}_2$ LGT, which discretizes the spatial coordinates $x_i=ia$ by introducing a lattice spacing $a$ and $L$ lattice sites, while retaining the continuity of time. $\mathbb{Z}_2$ charges are described by fermion creation and annihilation operators $c_i^\dagger,c_i^{\phantom{\dagger}}$ residing on the lattice sites, while the $\mathbb{Z}_2$ gauge field is defined on the lattice links $x_{i_\ell}=(i+\tfrac{1}{2})a$  through the Pauli operators $\{\sigma^x_{i_\ell, b}, \sigma^z_{i_\ell, b}\}$ labelled by the bond index $b\in\{1,\cdots,N_{\rm b}\}$.  In the following we work in  lattice units by setting $a=1$. Following the nomenclature  of multigraph theory, there is a total of $N_b$ so-called parallel edges connecting two matter sites, each of which is gauged with the corresponding gauge-field operator.  We note that all the results to be discussed below also apply to bosonic $\mathbb{Z}_2$ charges in the hardcore limit. 

Setting $\hbar = 1$, the Hamiltonian of this multi-link $\mathbb{Z}_2$ LGT depicted in Fig.~\ref{fig:z2_scheme}{\bf (a)} reads as follows
\begin{equation}
\label{eq:model}
    H=\frac{t}{2}\sum_{i,b}  \! \big(c_i^\dagger{\sigma_{i_\ell,b}^z} c_{i+1}^{\phantom{\dagger}} + {\rm H.c.}\big)  + \frac{  J}{2} \sum_{i,b}\sum_{a<b} \sigma^z_{i_\ell,a}{\sigma^z_{i_\ell,b}}+ \frac{h}{2}\sum_{i,b}{\sigma^x_{i_\ell,b}},
\end{equation}

\ni which can be shown to be invariant  under the local $\mathbb{Z}_2$ symmetry $[H,G_i]=0$. The symmetry generators are the following  involutory, $G_i^2=\mathbb{I}$,  and Hermitian, $G_i^\dagger=G_i$, generators 
\begin{equation}
\label{eq:generators}
    G_i = P_{i_\ell-1}^x (-1)^{n_i} P_{i_\ell}^x, \quad P^x_{i_\ell} = \prod_{b=1}^{N_b} \sigma_{i_\ell,b}^x, \hspace{1ex} n_i= c_{i}^{{\dagger}}c_{i}^{\phantom{\dagger}}.
\end{equation}
These operators, built from the fermion parity at a site and   parity-check stabilizers  of the link spins surrounding that site, are used to set the physical space by means of Gauss' law
\begin{equation}
\label{eq:gauss_law}
G_i\ket{\psi_{\rm phys}}=(-1)^{q_i}\ket{\psi_{\rm phys}},
\end{equation}
where the set $\mathbb{Z}_2$ background charges $q_i\in\{0,1\}$ fix the particular super-selection sector of the model. We work with an open chain: in the presence of dangling links, the gauge fields residing at the two ends of the chain never enter the dynamics of Eq.~\eqref{eq:model}. We can then fix them to fulfill $P^x_{1_\ell-1} = P^x_{L_\ell} = -1$ and consider an open chain terminating with sites. With this choice, the Gauss's law is deformed at the edges to $G_1 = -(-1)^{n_1} P_{1_\ell} = 1$ and $G_L = -P^x_{(L-1)_\ell} (-1)^{n_L} =1$.

In the above Hamiltonian,  $\sigma^z_{i_\ell,b}$ act as the parallel transporters ensuring  gauge invariance when the matter particles tunnel along any of the multiple links connecting two neighboring sites. The magnetic field term is described by the sum of all the possible minimal Wilson loops 
\begin{equation}
\label{eq:wilson}
W^{a,b}_{i_\ell,\circlearrowleft}=\sigma^z_{i_\ell,a}\sigma^z_{i_\ell,b},
\end{equation}
leading to a gauge-invariant
all-to-all interaction among the gauge-fields of strength  $J$, which aims at ordering the different flux configurations.   The remaining Pauli operators  $\sigma^x_{i_\ell,b}$ are used in an electric field term with strength  given by  $h$, introducing quantum fluctuations on the flux states (see Fig.~\ref{fig:z2_scheme}{\bf (b)}). We note that the standard $\mathbb{Z}_2$ gauge theory corresponds to $N_b=1$, while the $\mathbb{Z}_2$ loop-gauge theory studied in~\cite{domanti2025dynamical} requires setting $N_b=2$. In this paper, we showed how the phenomenon of Aharonov-Bohm (AB) caging~\cite{vidal1998aharonov} also appears for a  fully quantum dynamical gauge field, and how  it can lead to  a tight confinement of the $\mathbb{Z}_2$ charges into neutral dimers even for weak electric fields. In this article, we want to explore higher linking numbers $N_b$, particularly focusing on $N_b$ odd, such that this caging is no longer present as tunneling paths along different links can only yield pairwise destructive interference. Changing the linking number $N_b$ allows us to modify the lattice geometry, which is known to
have non trivial effects on the physics of gauge theories~\cite{Fradkin_2013}. Since caging and tight confinement is not expected, we want to explore how different collective phenomena arise. 

We note that the magnetic interactions resemble $L$ copies of the Lipkin-Meshkov-Glick (LMG) model of nuclear physics~\cite{lipkin1965validity}, where $L$ is the number of lattice sites. In the large-$N_{b}$ limit and for $h<h_c=JN_b$, this model is known to host a mean-field-type phase transition in which the spins of each individual link develop a non-vanishing order parameter $\langle \sigma_{i_\ell,b}^z\rangle$  breaking the global inversion symmetry $\sigma_{i_\ell,b}^z\mapsto-\sigma_{i_\ell,b}^z$ within each link~\cite{ribeiro2008exact}. This SSB breaking is, however, incompatible with  the  local  symmetry of our LGT when coupling the  spins to the fermionic matter. 
Here, we are interested in some other mechanism of SSB  that is not in conflict with Elitzur's theorem and, moreover, can also appear at finite $N_b$ and go beyond  a mean-field behavior.  The perspective of the LMG model is still useful, and we  can still borrow some features of its treatment, such as the use of collective spin operators for each link $\boldsymbol{S}_{i_\ell} = \tfrac{1}{2}\sum_{b=1}^{N_b} \boldsymbol{\sigma}_{i_\ell,b}$. Neglecting a constant term, Eq.~\eqref{eq:model} has a 
neat expression in this formulation  
\begin{equation}
\label{eq:totspin}
    H=\sum_i  t \big(c_i^\dagger S^z_{i_\ell}c_{i+1}^{\phantom{\dagger}} + {\rm H.c.}\big)  + h \sum_i S^x_{i_\ell} + J \sum_i(S_{i_\ell}^z)^2 \, .
\end{equation}

\ni  In addition to the  gauge symmetry which, in terms of the collective spin operators, reads  $G_i = -{\rm exp}\{{\rm i}\pi (S_{i_\ell-1}^x+n_i+ S_{i_\ell}^x)\}$, one can check that the quadratic Casimir  operator of the $\mathfrak{su}$(2) algebra per link is also  conserved $[H,\boldsymbol{S}_{i_\ell}^2]=0$. Hence,  we can work with any of the possible eigenvalues $s_{i_\ell}(s_{i_\ell}+1)$ with $s_{i_\ell}\in\{J_{\rm min}, J_{\rm min} + 1,\cdots,N_b/2\}$ and $J_{\rm min} = 0\,(1/2)$ if $N_b$ is even (odd).

We point out two facts: first, when $N_b$ is  very large, $J$ would need to be rescaled to $J/N_b$ to ensure that the total energy-per-spin does not diverge~\cite{ribeiro2008exact}. This in fact underlies the previous expression of the critical transverse field in the LMG model. Second, the eigenvalues $m_{i_\ell}$ of $S^z_{i_\ell}|m_{i_\ell}\rangle=m_{i_\ell}|m_{i_\ell}\rangle$ can only be zero if $N_b$ is even. In the aforementioned loop-chain with  $N_b=2$~\cite{domanti2025dynamical}, the states corresponding to $m_{i_\ell}=0$ were identified with $\pi$-flux visons yielding a destructive AB interference in the particle tunneling dynamics, and leading to the partitioning of the lattice into disconnected AB cages~\cite{domanti2025dynamical}. However, when $N_b$ is an odd number, $m_{i_\ell}\neq0$ forbids the phenomenon of AB caging, and a genuinely different  phenomenology arises. To investigate it, we resort to the local conservation of $\boldsymbol{S}^2_{i_\ell}$ and restrict our attention to the highest-weight states corresponding to $s_{i_\ell} = {N_b}/{2}$ - see Fig.\ref{fig:z2_scheme}{\bf (b)} for the three-link case $N_b=3$. 
We note that  this figure draws a naive picture in which  the  effective tunneling of the fermions under the different flux configurations  can have different strengths depending on the projection of the total spin, which already hints towards a plausible Peierls' mechanism of SSB. Note, however, that $S_{i_\ell}^z\mapsto -S_{i_\ell}^z$ is implemented by the local $\mathbb{Z}_2$  symmetry generators, such that states with non-zero values of $\langle S_{i_\ell}^z\rangle $ are not gauge invariant. Hence, this mean-field argument leading to different tunneling strengths is not valid, and making Peierls' SSB compatible with gauge symmetry is  more subtle.

To discuss gauge-invariant states, we use the common eigenbasis of $P^x_{i_\ell}={\rm i} \, {\rm exp}\{{\rm i}\pi S_{i_\ell}^x\}$ and  $(S^z_{i_\ell})^2=\sum_{a,b}\tfrac{1}{4}W^{a,b}_{i_\ell,\circlearrowleft}$, which thus encodes the parity checks and the simplified magnetic-type Wilson fluxes along all loops~\eqref{eq:wilson}. In light of $P^x_{i_\ell}S^z_{i_\ell}P^x_{i_\ell}=-S^z_{i_\ell}$, this basis always exists and reads 
\begin{equation}
|m_{_{i_\ell}};\pm\rangle = \tfrac{1}{\sqrt{2}}\big(|m_{i_\ell}\rangle \pm |-m_{i_\ell}\rangle\big),
\end{equation}
for  $m_{i_\ell} \in \big\{\tfrac{1}{2},\dots, \tfrac{N_b}{2}\big\}$ strictly positive. These states are depicted in Fig.~\ref{fig:z2_scheme} {\bf (c)} using fuzzy spheres, and  correspond to  the (anti)symmetric superposition of two spin states with opposite projected magnetizations. In the large-$N_b$ limit, they become a superposition of  mesoscopically distinct states~\cite{PhysRevLett.82.1835}, which are typically referred to as cat states. In our present context,  these states satisfy 
\begin{equation}
\label{eq:good_basis}
    P^x_{{i_\ell}} |m_{i_\ell};\pm\rangle = \pm |m_{i_\ell};\pm\rangle,\hspace{2ex}(S^z_{i_\ell})^2 |{m_{i_\ell};\pm}\rangle = m_{i_\ell}^2 |m_{i_\ell};\pm\rangle.
\end{equation}
Hence,  the $\pm$ labels denote the eigenvalues of the parity check, corresponding to the symmetric or antisymmetric nature of the superposition of the cat state. In the neutral gauge sector $q_i=0$ $\forall i$~\eqref{eq:gauss_law}, links that are adjacent to an occupied (empty) matter site correspond to spin states carrying opposite (equal) eigenvalues of the parity check (see Fig.~\ref{fig:z2_scheme} {\bf (d)}). 

We note that, in the standard $\mathbb{Z}_2$ chain, parity checks  contain a single link, such that  
$|m_{i_\ell};\pm\rangle$  have a fixed $m_{i_\ell}=1/2$ and are the eigenvectors of $\sigma^x_{i_\ell}$,   coinciding with the Hadamard electric-field basis $|\tfrac{1}{2};\pm_{i_\ell}\rangle = \tfrac{1}{\sqrt{2}}(|\uparrow_{i_\ell}\rangle \pm |{\downarrow_{i_\ell}}\rangle)=|\pm_{i_\ell}\rangle$.  In contrast, for our  multi-link $\mathbb{Z}_2$ gauge theory, the higher-spin representation introduces a much richer  structure: owing to the Gauss law constraints~\eqref{eq:gauss_law}, a given  arrangement of $\mathbb{Z}_2$ charges unequivocally determines the distribution of the link parity-check $P^x_{i_\ell}$  eigenvalues along the chain, but leaves the values of $m_{i_\ell}$ unconstrained. As such, contrasting  the situation in the standard $\mathbb{Z}_2$ LGT on a chain~\cite{borla2020confined}, it is not possible to solve for the Gauss constraints and rewrite the model purely in terms of  an effective  Hamiltonian that only involves the matter fields. The multi-link $\mathbb{Z}_2$ LGT is arguably the simplest lattice model where the gauge field dynamics is not entirely fixed by that of the matter fields, and this leads to  very  different physical phenomena, as we  now discuss.

For future convenience, we note that the $S=3/2$ operators can be written in the common eigenbasis $|m_{i_\ell};\pm\rangle$  by introducing  two sets $\boldsymbol{\rho}_{i_\ell},\boldsymbol{\tau}_{i_\ell}$ of Pauli operators  
\begin{equation}
\label{eq:tau_and_rho}
S^x_{i_\ell} = \tfrac{\sqrt{3}}{2} \tau^x_{i_\ell}\otimes\mathbb{I}_2+\tfrac{1}{2} \big(\mathbb{I}_2-\tau^z_{i_\ell}\big)\otimes\rho^z_{i_\ell},\hspace{2ex}S^z_{i_\ell} =  \big(\mathbb{I}_2+\tfrac{1}{2}\tau^z_{i_\ell}\big)\otimes\rho^x_{i_\ell}.
\end{equation}
 The  parity checks $P^x_{i_\ell} = \mathbb{I}_2\otimes\rho^z_{i_\ell}$ and the Wilson fluxes encoded in $(S^z_{i_\ell})^2 = (\tfrac{5}{4}\mathbb{I}_2+\tau_{i_\ell}^z)\otimes\mathbb{I}_2$ clearly commute with one another after this mapping, which also  shows  that the gauge-field  parity (flux) is entirely encoded in the $\rho$($\tau$) degrees of freedom. Such a decomposition follows from identifying the basis states as tensor products $|m_{i_\ell};\pm\rangle := | \tau_{i_\ell}\rangle\otimes|\rho_{i_\ell}\rangle$, with $\rho_{i_\ell}^z|\rho_{i_\ell}\rangle = \rho_{i_\ell} \ket{\rho_{i_\ell}}$ and $\tau^z_{i_\ell} |\tau_{i_\ell}\rangle = \tau_{i_\ell} \ket{\tau_{i_\ell}}$. In particular, we find  $\ket{3/2;\pm} = \ket{1}\otimes\ket{\pm 1}$ and $\ket{1/2; \pm} = \ket{-1}\otimes\ket{\pm 1}$ for the three-link case. The values of $\rho_{i_\ell}$ and $\tau_{i_\ell}$ respectively coincide with the eigenvalues of $P^x_{i_{\ell}}$ and 
 \begin{equation}
 \label{eq:T_opt}
 T_{i_\ell} = \big(S^z_{i_\ell}\big)^2-\tfrac{5}{4}\mathbb{I}_2=\sum_{a,b}\tfrac{1}{4}W^{a,b}_{i_\ell,\circlearrowleft}-\tfrac{5}{4}\mathbb{I}_2.
  \end{equation}
   Up to an irrelevant constant, the Hamiltonian of the model can thus be rewritten as
\begin{align}
\label{eq:gauged_z2bh}
\nonumber
    H = &\,t\sum_i \, c_i^\dagger\!\left(\big(1+\tfrac{1}{2}\tau_{i_\ell}^z\big) \otimes \rho_{i_\ell}^x \right)c_{i+1}^{\phantom{\dagger}} + {\rm H.c.} \,\\ 
    + &J\sum_i \tau_{i_\ell}^z+h\sum_i\Big(\tfrac{\sqrt{3}}{2} \, \tau^x_{i_\ell} + \tfrac{1}{2} \,  (1-\tau^z_{i_\ell})\otimes\rho^z_{i_\ell}\Big). 
\end{align}

\ni The gauge-symmetry generators can be represented solely in terms of the fermionic and $\rho$ degrees of freedom, namely
\begin{equation}
\label{eq:eff_gauge_symmetry}
G_i = \rho^z_{i_\ell-1} (-1)^{n_i} \rho^z_{i_\ell}.
\end{equation}
The tunneling term in Eq.~\eqref{eq:gauged_z2bh}, being proportional to $\rho^x_{i_\ell}$, inverts the parity check as the fermions tunnel, which in Fig.~\ref{fig:z2_scheme}{\bf (c)} amounts to changing the antisymmetric superposition of the cat state to a symmetric one $\ominus\mapsto\oplus$ upon tunneling.
Unless otherwise specified, we work in the neutral gauge sector $q_i=0$ $\forall i$~\eqref{eq:gauss_law}. Even if this formulation~\eqref{eq:gauged_z2bh} appears more convoluted than the original one based on the total spin~\eqref{eq:totspin}, we will now show that it is a key step to formalise the aforementioned Peierls' mechanism of SSB in the presence of gauge symmetry, which is the trigger for all the interesting topological effects, fractionalization and deconfinement that follow.

\begin{figure*}[th]
    \centering
    \includegraphics[width = 0.9 \linewidth]{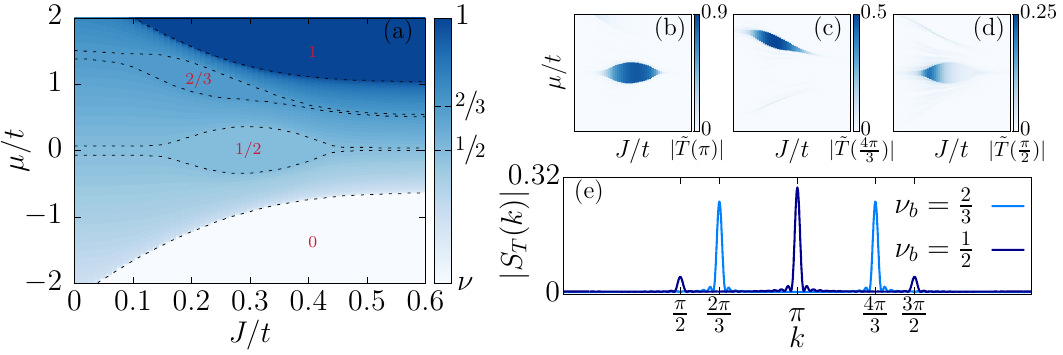}
    \caption{\textit{Incompressible regions and LRO at small $h/t$}. Panel (a) displays the filling fraction $\nu$ in the ground state of the system as a function of $J/t$ and $\mu/t$, at $h/t = 0.1$. Incompressible regions are stabilized in which the filling fraction is fixed to $0,1/2,2/3$ or $1$, as denoted by the red labels inside each lobe and are delimited by dashed lines. In the regions corresponding to $\nu = 1/2,2/3$ the lattice translational symmetry is spontaneously broken, as manifested by a spatially-periodic modulation of $\langle T_{i_\ell}\rangle$, whose Fourier transform $\tilde{T}(k)$ is peaked at the order wave-vector. Three different orderings appear: in the half-filled region, $\tilde{T}(k)$ has two peaks, a dominant one at the Fermi momentum $k_F(1/2)=\pi$ (b) and a sub-dominant one at $\tilde{k}=\frac{\pi}{2}$ (d); in the $\nu = 2/3$ lobe, $\langle T_{i_\ell} \rangle$ oscillates with a period of three lattice spacings, corresponding to a peak in the Fourier transform at the Fermi momentum $k_F(2/3)=\frac{4\pi}{3}$ (c) - see main text. In panel (e) we show that the structure factor associated to the correlator $C^T_{ij}$, $S_T(k)$, evaluated at $J/t = 0.2$ and $h/t = 0.1$ in the incompressible regions, is peaked at the order wave-vectors, a signature of long-range order - see main text. All the figures correspond to $L=120$ sites.}
    \label{fig:cp_analysis}
\end{figure*}
\end{section}

\begin{section}{\bf Peierls-type instability and    Wilson loop order}
\label{sec:peierls}

Let us momentarily switch off the quantum fluctuations induced by the electric field $h=0$. In this limit,  the gauge-invariant tunneling and magnetic-type term of the Hamiltonian mutually commute~\eqref{eq:totspin}. In particular, the operators $(S^z_{i_\ell})^2$ are locally conserved, and the physical Hilbert space thus decouples into many disconnected sectors identified by all the possible choices of the eigenvalues $m^2_{i_\ell}$ on the bonds of the chain. Using the formulation in Eq.~\eqref{eq:gauged_z2bh} we can  show that, in any of these sectors, the system can be reduced to a tight-binding model. This is achieved by introducing the following  gauge-invariant  fermion operators  
\begin{equation}
 \label{eq:dressed_fermions}
d_i = \prod_{j
< i} \rho^x_{j_\ell} \, c_i,
\end{equation}
which are dressed by a parity-conjugate string operator  connecting the left boundary to the fermion site to ensure for gauge  invariance under~\eqref{eq:eff_gauge_symmetry}. Working in the different flux sectors given by the $\tau$ basis,  the Hamiltonian reduces to
\begin{equation}
\label{eq:tight_binding}
    \tilde{H} = \sum_{i,\tau_{i_\ell}}  \!\Big(t(\tau_{i_\ell})d_i^\dagger d^{\phantom{\dagger}}_{i+1}  + \tfrac{1}{2}J \tau_{i_\ell}\Big)\!\ket{\tau_{i_\ell}}\!\bra{\tau_{i_\ell}} +{\rm H.c.},
\end{equation} 

\ni where we have introduced the flux-dependent  tunneling amplitudes 
\begin{equation}
    t({\tau}_{i_\ell})=\left(1+\tfrac{1}{2}\tau_{i_\ell}\right)t,
\end{equation} where $\tau_{i_\ell} = (m_{i_\ell}^2 - 5/4) \in \{-1,1\}$ denotes the eigenvalue of $T_{i_\ell}$~\eqref{eq:T_opt}. We have thus found a gauge-invariant mechanism through which the strength of the tunnelings can actually become inhomogeneous, i.e. $t_1=t(-1)=t/2$ and $t_2=t(+1)=3t/2$ for the three-link case, being the second tunneling  depicted schematically in Fig.~\ref{fig:z2_scheme}{\bf (c)}.

The ground-state  can  be determined by finding the optimal configuration $\{\tau_{i_\ell}\}$ that minimizes the energy: while the tunneling term would be minimized by a homogeneous background $\tau_{i_\ell} = 1$ $
\forall i$, this configuration is penalized by the magnetic-type term when $J>0$, which favors states in which some of the  spherical caps are pierced by $\pi$ fluxes, requiring an inhomogeneous arrangement of the spins (see Fig.~\ref{fig:z2_scheme} {\bf (b)}). We anticipate that the competition of these two mechanisms is at the roots of a SSB analogous to the Peierls' instability in 1D metals~\cite{peierls2001quantum}, where it is energetically favorable to dimerize the lattice  by opening an energy gap at the Fermi surface. From this perspective,  $\tau_{i_\ell}$ could be understood as a discretized version of the   elastic distortion of a crystal modulating the tunneling as in electron-phonon lattice models~\cite{PhysRevLett.42.1698}. However, note that the competition in our model is not between the elastic deformation energy and that of a Fermi liquid, but between the magnetic energy of a configuration of gauge fluxes in our multi-link $\mathbb{Z}_2$ LGT and the associated kinetic energy of the dynamical $\mathbb{Z}_2$ charges. Depending on the filling fraction $\nu = \frac{N_{\rm f}}{L}$, where $N_{\rm f}$ is the number of fermionic charges in the system, and on the  specific ratio of the parameters $J/t$, one can find different inhomogeneous orderings of $\tau_{i_\ell}$. Some of this open up a gap in the single-particle energy spectrum at the Fermi momentum $k_F = 2 \pi \nu$, in  analogy to the Peierls' mechanism. 

As we will show later on, the Peierls-type SSB in this specific limit can  be solved analytically. However, before describing this solution, it is instructive to realise that the  
effective model~\eqref{eq:gauged_z2bh} prior to the dressing in Eq.~\eqref{eq:dressed_fermions} can  be seen as a gauge-invariant version of  the $\mathbb{Z}_2$ Bose-Hubbard model~\cite{gonzalezcuadra2019symmetry,gonzalezcuadra2020zn} in the  hardcore-boson limit. In that model,  $\tau^z_{i_\ell}, \tau^x_{i_\ell}$ play the role of discretised ``phonon" modes with canonical coordinates $q_{i_\ell},p_{i_\ell}$, respectively. Note that $\mathbb{Z}_2$ refers to the two-level nature of these ``phonons", and not to any strict local gauge symmetry~\eqref{eq:eff_gauge_symmetry}. In Eq.~\eqref{eq:gauged_z2bh} the additional $\rho^x_{i_\ell},\rho^z_{i_\ell}$ operators play a key role in  promoting the global  parity symmetry of the $\mathbb{Z}_2$ Bose-Hubbard model to a local $\mathbb{Z}_2$ gauge symmetry. When switching back the  electric field term $h>0$, we see from Eq.~\eqref{eq:gauged_z2bh} that its effect can be split into two contributions: the first introduces quantum fluctuations affecting  the magnetic fluxes ($\tau$), mimicking the competition of  the kinetic and deformation energy in the Peierls' mechanism. In contrast, the second term is a  new term that activates  quantum fluctuations in the gauge parity ($\rho$) leading, as we show below,   to an effective  string tension as that observed in  $\mathbb{Z}_2$ LGTs~\cite{borla2020confined,kebric2021confinement}. This string tension is  actually conditioned on the flux: it only acts on
 states $\ket{-1,\pm 1}$ with  different configurations of $\pi$ fluxes threading two out of the three spherical caps - see Fig.~\ref{fig:z2_scheme}.

When a small electric field is turned on $h\neq 0$, quantum fluctuations are activated that disrupt the local conservation of $T_{i_\ell}$~\eqref{eq:T_opt}, and the simple tight-binding model with dressed fermions~\eqref{eq:tight_binding} is superseded by Eq.~\eqref{eq:gauged_z2bh}.  To address this regime, we resort to matrix product state (MPS) methods using a variational scheme with sequential optimizations based on the density matrix renormalization group (DMRG) scheme of sweeps~\cite{schollwock2011the,fishman2022the,itensor-r0.3}. To account for the dependence of the ordered phases on the fermion filling $\nu$,  we introduce a chemical potential term $H\mapsto H({\mu})=H-\mu \sum_i n_i$ into Eq.~\eqref{eq:totspin}. This allows us to explore different   total numbers of $\mathbb{Z}_2$ charges   and, thus, different fermion fillings in the system by calculating the MPS state that is the closest to the groundstate of the grand-canonical Hamiltonian 
\begin{equation}
\label{eq:mps}
\ket{g}={\rm argmin}\{\bra{\psi_{\rm MPS}}H({\mu}) \ket{\psi_{\rm MPS}}\}.
\end{equation}
To detect the onset of the Peierls-type inhomogeneous phases, we analyze the spatial profile of the magnetic-flux operator $\langle T_{i_\ell}\rangle$ as a function of the bond-index $i_\ell$ along the chain. The results for $h/t = 0.1$ are shown in Fig.~\ref{fig:cp_analysis}, where we plot the filling fraction $\nu=-\frac{1}{L}\partial_\mu\bra{g}H(\mu)\ket{g}$  as a function of $J/t$ and $\mu/t$. We find wide incompressible regions $\kappa=\partial_\mu\nu=0$ that correspond to fixed filling fractions, the largest ones relative to $\nu \in\{ 0,\frac{1}{2},\frac{2}{3},1\}$. Let us now discuss the underlying Wilson loop order in these incompressible phases.

The regions with filling $\nu=0,1$ can be studied analytically by diagonalizing the local pure-gauge part of the Hamiltonian~\eqref{eq:gauged_z2bh}, as the tunneling term vanishes  on the empty and fully-filled regimes. In these cases, the parity operators $P^x_{i_\ell}$ are conserved and the eigenvalues $\rho_{i_\ell}$ at every bond are fixed by  Gauss' law. The eigenstates of the  Hamiltonian in these regimes  can be written as 
\begin{equation}
\label{eq:states}
    \ket{\epsilon^\alpha_{i_\ell}(\rho_{i_\ell})} = \sum_{\tau=\pm1} c^\alpha_\tau(\rho_{i_\ell}) \, \ket{\tau_{i_\ell}} \otimes \ket{\rho_{i_{\ell}}},
    \end{equation}
    where $\alpha=\pm$ labels the two eigenvalues $\epsilon^-_{i_\ell}(\rho_{i_\ell}) < \epsilon^+_{i_\ell}(\rho_{i_\ell})$ for $\rho_{i_\ell} = \pm 1$ of the pure gauge-field operator 
    \begin{equation}
        O_{i_\ell}=J  \tau_{i_\ell}^z+h \Big(\tfrac{\sqrt{3}}{2} \, \tau^x_{i_\ell} + \tfrac{1}{2} \,  (1-\tau^z_{i_\ell}) \rho_{i_\ell}\Big).
    \end{equation}
    Diagonalizing this matrix, one obtains the eigenvalues 
    \begin{equation}
    \epsilon_{i_\ell}^\pm(\rho_{i_\ell}) = (\tfrac{5}{4} J +\tfrac{1}{2} h\rho_{i_\ell} ) \pm \sqrt{J^2 -  Jh\rho_{i_\ell} + h^2},
    \end{equation}
    and the associated eigenvectors $c^\alpha_\tau(\rho_{i_\ell})$ used in Eq.~\eqref{eq:states}, which have a well-defined  parity $P^x_{i_\ell} |\epsilon^\alpha_{i_\ell}(\rho_{i_\ell})\rangle = \rho_{i_\ell} |\epsilon^\alpha_{i_\ell}(\rho_{i_\ell})\rangle$.  The exact ground state for $\nu=0,1$ is obtained by  taking the tensor product of all the local lowest-energy states
    \begin{equation}
    \ket{g(\rho_{i_\ell})} = \bigotimes_i |\epsilon_{i_\ell}^-(\rho_{i_\ell})\rangle,
       \end{equation}
       such  that  we  find the Wilson loops~\eqref{eq:T_opt} arrange  as follows 
       \begin{equation}
       \label{eq:Wilson-loop_order}
           T(\rho_{i_\ell}):=\langle g(\rho_{i_\ell}) \vert T_{i_\ell} \vert g(\rho_{i_\ell}) \rangle =\frac{  \rho_{i_\ell} h- 2J}{2 \sqrt{J^2 -  Jh\rho_{i_\ell} + h^2}}.
       \end{equation}
       
       For $\nu = 0$,  Gauss' law determines an homogeneous ordering of the parity checks $\rho_{i_\ell} = \rho$ $\forall i$, which induces a translationally-invariant arrangement of the fluxes $\langle T_{i_\ell} \rangle = T(\rho)$ $\forall i$. We note that only the choice $\rho=-1$ is compatible with the deformation of the Gauss law at the edges $G_1 = -(-1)^{n_1} P^x_{1_\ell}=1$, $G_L = -P^x_{(L-1)_\ell} (-1)^{n_L} = 1$. In contrast, at  filling $\nu=1$,  Gauss' law requires an alternation of $\rho_{i_\ell}= \pm 1$ that leads to  an inhomogeneous dimerized flux pattern  $\langle T_{i_\ell} \rangle = T(1) \, \delta_{i,{\rm odd}} + T(-1) \, \delta_{i,{\rm even}}$. We note that the other alternating configuration, obtained by a global parity flip $\rho_{i_\ell} \to -\rho_{i_\ell}$, is not a physical state satisfying the  Gauss law. In this case, the dimerization is unique and is due to the aforementioned flux-dependent string tension term of Eq.~\eqref{eq:gauged_z2bh}, which vanishes for $h = 0$. In contrast, for a system with periodic boundary conditions (PBC), both  inhomogeneous patterns carrying opposite dimerizations are allowed, leading to a two-fold groundstate degeneracy. In any case,  no breakdown of the lattice translational invariance actually occurs, as the ground state will be an arbitrary combination of these two degenerate configurations. 

The situation is very different  for fractional fillings $\nu = \frac{1}{2},\frac{3}{2}$, where the tunneling of the fermions no longer vanishes, and the Wilson loop order can no longer be calculated exactly~\eqref{eq:Wilson-loop_order}. By resorting to MPS~\eqref{eq:mps}, we find  an inhomogeneous Wilson loop order, which is associated to the spontaneous breakdown of the lattice translational symmetry and to a specific distribution of the fluxes. This SSB is identified by inspecting the spatial modulation of $\langle T_{i_\ell} \rangle$ via its Fourier series 
\begin{equation}
    \tilde{T}(k)=\frac{1}{N_s
} \sum_{j} {\rm e}^{{\rm i} k a j} \langle g| T_{j_\ell}|g\rangle,
\end{equation}
where $N_s = 3L/5$ is the number of sampled sites and the sum runs over the values of $j \in \{\frac{L}{5}+1, \dots, \frac{4L}{5} \}$. We work with system sizes $L$ such that the $N_s$ sampled sites contain an integer number of unit cells in any of the observed inhomogeneous orderings. With such prescription, which is carried on to any of the observables discussed below, we aim at reducing finite-size effects arising from getting too close to the edges of the system. In Figs.~\ref{fig:cp_analysis}{\bf (b)}-{\bf (d)}, we see that the maximum 
\begin{equation}
    \tilde{T}_{\rm peak}={\rm max}\{\abs{\tilde{T}(k)}: k\in{\rm BZ}\}
\end{equation}
plays the role of an order parameter sensitive to the different inhomogeneous orderings.
We find that the momentum of the peak coincides with the Fermi wave-vector $ k_F = 2\pi\nu$, and  leads to  oscillations of period of 2 ($\nu=\frac{1}{2}$) or 3 ($\nu = \frac{2}{3}$). 

Let us now discuss this SSB in more detail. The lattice translational symmetry in the formulation~\eqref{eq:totspin}  is captured by the cyclic group $\mathsf{G}=\{\mathcal{T}^n\,:n\in\{1,\cdots,L\}\}\simeq\mathbb{Z}_L$, where $\mathcal{T}c_i\mathcal{T}^{-1}=c_{i+1}$, and  $\mathcal{T}\boldsymbol{S}_{i_\ell}\mathcal{T}^{-1}=\boldsymbol{S}_{(i+1)_\ell}$, and we have assumed periodic boundary conditions $\mathcal{T}^L=\mathbb{I}$.  This symmetry is spontaneously broken down to the subgroups $\mathsf{H}_\ell=\{\mathcal{T}^{\ell n}\,:n\in\{1,\cdots,L/\ell\}\}\simeq\mathbb{Z}_{L/\ell}$, which describe  translations of a larger unit cell of size $\ell$, and we assume that the chain always holds an integer number of those. While $\ell=3$ at the filling $\nu = 2/3$, for $\nu=1/2$ a sub-modulation appears, together with the leading period-2 oscillation, associated with the wave-vector $\tilde{k}=\pi/2$. This results in the breakdown of the lattice translational invariance into its subgroup $\mathbb{Z}_{L/4}$, effectively enlarging the unit cell to $\ell=4$ sites. As a consequence, the groundstate shows a degeneracy characterized by the corresponding cosets $\mathsf{G}/\mathsf{H}_\ell=\mathbb{Z}_\ell$, which describe the different ways in which the inhomogeneous flux distributions can be arranged within the corresponding unit cells. SSB
manifests in long-range order (LRO), which can be inferred by the asymptotic behavior of the correlation function 
\begin{equation}
    C^T_{ij}=\langle g| (T_{i_\ell} - \overline{T})( T_{j_\ell} - \overline{T})|g\rangle,
\end{equation}
where $\overline{T} = \frac{1}{N_s} \sum_i \langle g| T_{i_\ell}|g\rangle$ - regardless of the ground state degeneracy.  In Fig.~\ref{fig:cp_analysis}, we   present the  structure factor 
\begin{equation}
    S_T(k) = \frac{1}{N_s^2} \sum_{i,j} {\rm e}^{{\rm i} k a (i-j)} C^T_{ij},
\end{equation} 
obtained with our MPS approach~\eqref{eq:mps}.
This figure clearly shows that, in the incompressible regions, the structure factor has a dominant peak at the predicted  Fermi wave-vector $k_F\in\{\pi,\tfrac{2\pi}{3}\}$ for $\nu\in\{\tfrac{1}{2},\tfrac{3}{2}\}$.

\end{section}

\begin{section}{\bf Peierls-type instability and topological bond-order waves}
\label{sec:tbow}

We have so far focused on pure-gauge observables, showing how the Peierls' mechanism for the SSB of translational symmetry manifests through different periodic arrangement of Wilson fluxes. However, the distribution of $\mathbb{Z}_2$ charges is deeply intertwined with the gauge fields in constrained theories such as the present LGT~\eqref{eq:gauss_law}. As we now show, the periodic modulation of $T_{i_\ell}$  is accompanied by  periodic spatial oscillations of the gauge-invariant bond operator 
\begin{equation} 
\label{eq:gauge_inv_bond}
B^{\phantom{\dagger}}_{i,i+1} = c^\dagger_i S_{i_\ell}^z c_{i+1}^{\phantom{\dagger}} + {\rm H.c.}.
\end{equation} 
\begin{figure}
    \centering
    \includegraphics[width=0.9\linewidth]{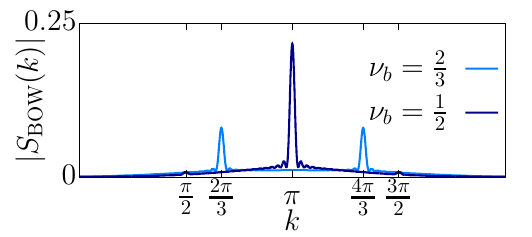}
    \caption{\textit{BOW structure factor.} The BOW structure factor $S_{\rm BOW}(k)$ is displayed for $\nu = \frac{1}{2}$ and $\nu=\frac{2}{3}$ fillings, for a chain of $L=120$ sites and $J/t = 0.2$, $h/t = 0.1$. In both cases, the behaviour of the BOW structure factor parallels that of $S_T(k)$ - see panel (e) of Fig.~\ref{fig:cp_analysis}, being peaked at the Fermi momentum $k_F=2\pi\nu$ and manifesting LRO. For the half-filled case, the subleading peak at the wave-vector $\tilde{k} = \frac{\pi}{2}$ is here smaller and is partially hidden by the curve at $\frac{2}{3}$ filling.}
    \label{fig:bow}
\end{figure}
By expressing this bond operator in the $\tau$-$\rho$ basis
\begin{equation} 
\label{eq:gauge_inv_bond}
B^{\phantom{\dagger}}_{i,i+1} =c^\dagger_i \Big(\big(\mathbb{I}_2+\tfrac{1}{2}\tau^z_{i_\ell}\big)\otimes\rho^x_{i_\ell}\Big) c_{i+1}^{\phantom{\dagger}} + {\rm H.c.},
\end{equation}
it becomes clear that both  the flux and parity degrees of freedom of the gauge field will intertwine with the density distribution of the fermionic $\mathbb{Z}_2$ charges when  translational symmetry is spontaneously broken. In fact, we find numerically using MPS methods that the above LRO is also manifested in a bond-ordered wave (BOW) captured by the structure factor 
\begin{equation}
S_{\rm{BOW}}(k) = \frac{1}{N_s^2} \sum_{i,j} {\rm e}^{{\rm i} ka (i-j)} \langle g|(B_{i,i+1} - \overline{B}) (B_{j,j+1} - \overline{B}) |g\rangle,
\end{equation}
where $\overline{B} = \frac{1}{N_s} \sum_i \langle g|B_{i,i+1}|g\rangle$. This structure factor manifests an analogous behaviour to $S_T(k)$, except for a reduced value of the subleading peak at $\nu = \frac{1}{2}$ - see Fig.~\ref{fig:bow}. In contrast to the standard bond- or charge-ordered waves in condensed matter~\cite{kumar2010bond,segupta2002bond,loida2017probing}, we note that the bare bond density $\mathcal{B}_{i,i+1}^{\phantom{\dagger}}=\langle g|(c^\dagger_i  c_{i+1}^{\phantom{\dagger}}+{\rm H.c.})|g\rangle$ cannot attain a non-zero value, as this would break the local gauge invariance. Interestingly, the gauge symmetry yields an obstruction to this bare LRO, such that the BOW needs to involve both the fermion operators and the link gauge fields, becoming a weight-4 operator~\eqref{eq:gauge_inv_bond}. This discussion parallels the gauge obstruction to the mean-field Peierls' argument $t\mapsto t\langle S_{i_\ell}^z\rangle$, which required a more careful discussion of the dependence of the effective tunneling strengths on the flux. 

It is interesting to note that for certain SSB groundstates, the spatial modulation of the effective tunneling strengths is reminiscent to that found in the semi-classical limit of the Su-Schrieffer-Heeger (SSH) model and variants thereof~\cite{huckel1931quant,su1979solitons}, which are archetypes of the physics of topological insulators and symmetry-protected topological phases of matter~\cite{kane2005z2,kane2005quantum,bernevig2013topological,hohenadler2013correlation,asboth2016lecture,rachel2018interacting}. As we show in the following section, this connection can be formalized and made compatible with the gauge symmetry.
To understand the role of topology in the BOW phase, we now focus on the half-filled regime in which  $\langle T_{i_\ell} \rangle$ displays spatial oscillations with two characteristic scales, one given by the expected Fermi wavevector $k_F = \pi$ manifest in the largest peak of the structure factor of Fig.~\ref{fig:cp_analysis}, and a faster modulation with $\tilde{k} = \frac{\pi}{2}$ leading to a smaller but differentiated peak. This is generated by the quantum fluctuations induced by a small electric field $h/t$, which vanishes at $h=0$.

\begin{subsection}{Symmetry-protected topology  at $h=0$}
We start this analysis  at $h=0$, where we can actually prove things analytically by leveraging the commutation $[T_{i_\ell},H] = 0$. This implies that the ground state can be written as a product state $\ket{\Psi} = \ket{\psi_m}\otimes\ket{\psi_\tau}$, where $\ket{\psi_m}$ is a gauge-invariant  state fulfilling $G_i \ket{\psi_m} = \ket{\psi_m}$. The  positions of the fermionic $\mathbb{Z}_2$ charges in this state  fully determine the configuration of the gauge-field parities $P^x_{i_\ell}$, while $\ket{\psi_\tau}$ fixes the flux configuration of $T_{i_\ell}$ on all the bonds of the chain. In this regime,  the Hamiltonian can be rewritten  in the dressed fermion basis~\eqref{eq:tight_binding}, such that the $N_{\rm f}$ fermions fill in different orbitals leading to  $\ket{\psi_m} = 
\sum_{\boldsymbol{i}}c_{i_1,\dots,i_{N_{\rm f}}} \bigotimes_{m=1}^{N_f} d^\dagger_{i_m} \ket{{\rm vac}}$, with $\ket{{\rm vac}}$ the gauge-invariant vacuum state. In the original picture, it consists of empty fermion sites and either of the two possible ferromagnetic orderings of $\rho^z_{i_\ell}$ on the bonds of the chain. In the dressed fermion picture, these  parity  orderings become invisible, and one works with a simple fermionic vacuum. On the other hand, the flux information of the gauge fields is encoded in $\ket{\psi_\tau} = \bigotimes_i \sum_{\tau=\pm1} d_{i_\ell,\tau} \ket{\tau_{i_\ell}}$, with $d_{i_\ell,\tau} \in \{0,1\}$ as the states cannot have coherences in the flux basis for $h=0$. 

A homogeneous ordering of $\langle T_{i_\ell} \rangle = \tau = \pm 1$ results into a gapless metallic phase. In the ground state, fermions occupy all the negative-energy single-particle states of the band $\epsilon(k,\tau) =-2t(1+\frac{\tau}{2})  \cos(k)$, $k \in [-\pi,\pi)$, leading to the ground state energy density 
\begin{equation}
    \label{eq:homogeneous_energy}
    \frac{E_g(\tau)}{L} = 
    -\frac{2t }{\pi}(1+\tfrac{\tau}{2})+ J\left(\tau + \tfrac{5}{4}\right) ,
\end{equation}
obtained by adding all the energies up to half filling, taking a continuum limit, and   also considering  the magnetic flux contribution. For the dimerized flux  pattern 
\begin{equation}
\langle T_{i_\ell}\rangle = \frac{(1-(-1)^i)}{2} \tau + \frac{(1+(-1)^i)}{2} \overline{\tau}
\end{equation}
the effective tunneling strengths of the fermions  depend on the specific values of $\tau$ and $\bar{\tau}$. Introducing $t_1=(1+\frac{\tau}{2}) \, t$, $t_2=(1+\frac{\overline{\tau}}{2})\, t$, we find 
\begin{equation}
\label{dimer_tbm}
  H_m =\langle\psi_\tau|H|\psi_\tau\rangle= \sum_{n=1}^{L/2} (t_1 \, d_{2n-1}^\dagger d_{2n}^{\phantom{\dagger}} + t_2 \, d^\dagger_{2n} d_{2n+1}^{\phantom{\dagger}} + {\rm H.c.}),
\end{equation}
a dimerized tight-binding  model. A similar semi-classical model was 
introduced to describe the physics of certain polymers, like polyacetylene~\cite{barford2013electronic,huckel1931quant,su1979solitons}, in which electron-phonon interactions lead to a Peierls instability, and to the deformation of the lattice with a doubled unit cell.

In our case, $t_1/t$ and $t_2/t$ can take either of the two values $1/2$ and $3/2$, which we will respectively refer to as weak (W) and strong (S) bonds, depending on the flux pattern. A two-band spectrum arises from the dimerization $\epsilon_{\pm}(k,\tau,\bar{\tau}) = \pm \sqrt{t_1^2(\tau) + t_2^2(\bar{\tau}) +2t_1(\tau)t_2(\bar{\tau})\cos(k)}$ in a reduced Brillouin zone $k\in\left[-\frac{\pi}{2},\frac{\pi}{2} \right)$. In the ground state at $\nu=\frac{1}{2}$ particles fully occupy the lower band ($-$) and leave the upper one empty. Its energy can be expressed in terms of the complete elliptic integral of the second kind $\mathsf{E}(x) = \int_0^\frac{\pi}{2} {\rm d}k [1-x\sin^2(k)]^{-\frac{1}{2}}$, as
\begin{align}
    \label{eq:dimerized_energy}
    \frac{E_g(\tau,\bar{\tau})}{L} &= -\frac{t_1+t_2}{\pi} \mathsf{E}\left(\frac{4t_1t_2}{(t_1+t_2)^2}\right) + J\frac{t_1^2+t_2^2}{2t^2},
\end{align}
which reads  $E_g(\tau,\bar{\tau})/L=-2t \mathsf{E}({3}/{4})/\pi + 5J/4$ for our specific values of the weak and strong bonds.

The competition of the magnetic flux and the  fermionic energies leads to a neat  gauge-invariant analogue of the $\mathsf{BDI}$-class topological insulator, a gapped system exhibiting a non-vanishing bulk topological invariant and symmetry-protected topological edge states~\cite{chiu2016classification,asboth2016lecture}. The trivial phase ($t_1 > t_2$, or S-W bond alternation) cannot be adiabatically connected  to the topological one ($t_1 < t_2$ or W-S bond alternation) without closing the gap at $t_1=t_2$. The topological protection is provided by the chiral (or sublattice) symmetry~\cite{bernevig2013topological}. These phases are distinguished by a different quantized value of a bulk topological invariant, the so-called Zak-phase~\cite{zak1989berry} 
\begin{equation}
\label{eq:zak}
\gamma = \oint_{BZ} \frac{{\rm d}k}{2\pi} \, \langle u_k|\partial_k|u_k\rangle.
\end{equation}
This is defined modulo $2\pi$ and is calculated as the lattice-momentum $k$ goes around the Brillouin zone (BZ) using the corresponding Bloch states$
\ket{u_k}$ of the filled energy band of Eq.~\eqref{dimer_tbm}. The trivial (topological) phase corresponds to $\gamma = 0$ ($\pi$) or, in terms of the winding number $\gamma = W \pi \, {\rm mod} \, 2\pi$, to $W = 0$ ($1$)~\cite{chiu2016classification,asboth2016lecture}.

By comparing the ground state energies of the ferromagnetic~\eqref{eq:homogeneous_energy} and dimerized~\eqref{eq:dimerized_energy} orderings, we find that
the two antiferromagnetic orderings $\tau = -\overline{\tau} = 1$ and $\tau=-\overline{\tau}=-1$, corresponding respectively to S-W and W-S alternations, are degenerate and energetically favourable in the range 
\begin{equation}
\label{eq:fo_crit}
J_{c,1}:= \frac{t(3-2E(\frac{3}{4}))}{\pi}< J < \frac{t(2E(\frac{3}{4})-1)}{\pi}=:J_{c,2} \, .
\end{equation}

\ni Hence, the lattice translational invariance is broken between these two critical values, separating the symmetry protected insulating phase from trivial metallic phases: for $J<J_{c,1}$ $\tau = \overline{\tau} = 1$, while for $J>J_{c,2}$ $\tau = \overline{\tau} = -1$.

In this regime of vanishing electric field, the  trivial and topological phases can arise  in the $\mathsf{G}/\mathsf{H}_2=\mathbb{Z}_2$ degenerate ground states associated to the two different Neel alternations of the gauge fluxes  $\langle T_{i_\ell}\rangle$. We point out that, for a finite-size open chain, the S-W ordering is always at a slightly lower energy than the W-S, effectively making the topological SPT phase an excitation. However, it is easy to foresee a straightforward  energetic penalty, for instance by the addition of a tiny staggered term $\lambda \sum_i (-1)^i \Delta_{i_\ell}$, with $\lambda > 0$, such that the WS ordering arises in the groundstate, and the fermionic sector develops topological edge states and a non-zero Zak phase. 

We point out that, for $h=0$, the transitions between the homogeneous and inhomogeneous orderings are  first order.
However, we shall see that a small electric field $h \neq 0$ activating quantum fluctuations of the fluxes $\langle T_{i_\ell} \rangle$ leads to the genuine SSB of translational invariance and results in second order phase transitions. To the best of our knowledge, this is the first time in which a symmetry-breaking topological insulator~\cite{kourtis2014combined}, i.e. a topological phase that is protected by a global symmetry, here chiral and inversion symmetries,  but also triggered by the SSB of another global symmetry, here translational invariance,  appears in a  LGT with a local gauge symmetry.
We thus have an exotic topological BOW phase in which the gauge-field fluxes and parities intertwine with the bond-ordered densities in the bulk and the topological edge states, marrying the various roles of global and local symmetries and topological matter of contemporary theoretical physics in a simple yet rich gauge model.  

\end{subsection}

\begin{subsection}{Symmetry-protected topology at $h\neq 0$}
\begin{figure}
    \centering
    \includegraphics[width=\linewidth]{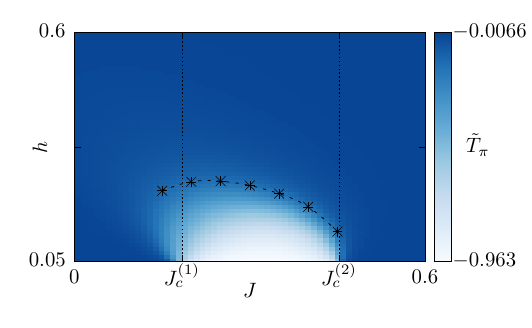}
    \caption{\textit{Dimerization order parameter.} We show the order parameter $\tilde{T}(\pi) = \frac{1}{N_s}\sum_i (-1)^i \, \langle T_{i_\ell} \rangle$ characterizing the long-range inhomogeneous flux ordering, for the system size $L=120$. The points $J_{c,1},J_{c,2}$, marked by the vertical dashed lines, are the critical points of a first-order transition and are reported in Eq.~\eqref{eq:fo_crit}, derived analytically at $h=0$. The black points that delimit the boundaries of the dimerized region are instead critical points of second order phase-transitions, found by the finite-scaling analysis reported at the end of Sec.~\ref{sec:tbow} and the dashed black curve connecting them has been drawn to lead the eye.}
    \label{fig:dimerization}
\end{figure}

\begin{figure*}
    \centering
    \includegraphics[width=0.9\linewidth]{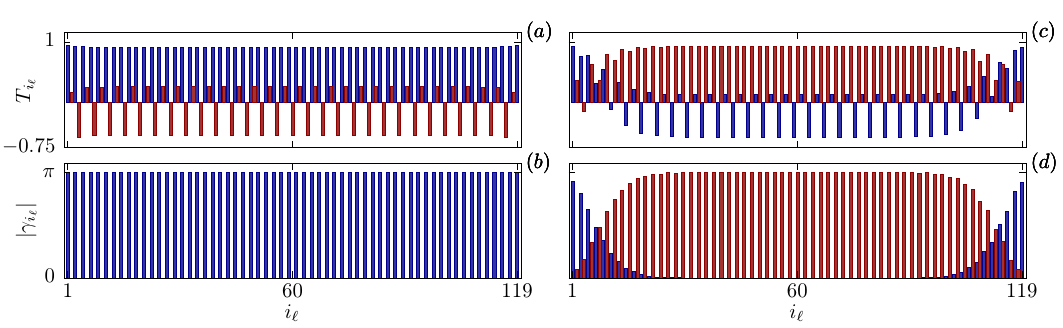}
    \caption{\textit{Local Berry phase}. We study the local Berry phase obtained by introducing a local twist in the hopping amplitude accross link $i_\ell$: $t_{i_\ell} \to {\rm e}^{\rm i \theta} t_{i_\ell}$ and considering a cyclic variation of $\theta$ - see text. In panel (a) we show the symmetry broken configuration of $T_{i_\ell}$ stabilized in the ground state of an open chain of $L=120$ sites at $J/t = 0.2$ and $h/t=0.1$, which alternates strong to weak bonds in a tetramerized pattern $S \, W_1 \, S \, W_2$, evidenced by a different box coloring at odd (blue) and even (red) bonds. In panel (b) we plot the absolute value of the local Berry phase obtained by twisting the corresponding bonds, resulting in the repeated pattern $\pi \, 0 \, \pi \, 0$. Panels (c) and (d) respectively show the inhomogeneous patterns of $\langle T_{i_\ell}\rangle$ and $\abs{\gamma_{i_\ell}}$ when the pinning potential~\eqref{eq:pinning} $V(\lambda = 5 \cdot 10^{-3} \, t)$ is applied to select a topological configuration, demonstrating localized domain walls whose spread is controlled by the value of $h/t$. The Berry phase's pattern $0 \, \pi \, 0 \, \pi$ is realized in the bulk of the chain, while its value is not quantized in the proximity of the localized edge modes.}
    \label{fig:local_berry}
\end{figure*}

Quantum fluctuations on the fluxes $\langle T_{i_\ell} \rangle$, which are detrimental to the stability of the inhomogeneous orderings, are activated as soon as a non-zero value of $h/t$ is turned on. Moreover, as already announced below Eq.~\eqref{eq:gauged_z2bh}, a non-zero $h/t$ supports a string-tension term, which depends on the state of $T_{i_\ell}$, and will induce non-local interactions between particles. Such string tension terms determine the onset of confining interactions between pairs of particles in the standard $\mathbb{Z}_2$ chain~\cite{borla2020confined,kebric2021confinement}: in the large $h/t$ limit, these pairs will be connected by short electric-field strings, forming neutral dimers as a result that can be understood as a cartoon verison of mesons. This tight confinement also appears  for $N_b=2$  in our model even when the electric field is not strong~\cite{domanti2025dynamical}. 
 We now want to explore the role of the quantum fluctuations for $N_b=3$,
 and characterize the effects of a small $h/t \neq 0$ on the phase diagram  at $\nu=\frac{1}{2}$, studying the fate of   the topological BOWs and connecting  to the physics of confinement/deconfinement. 
 
 In Fig.~\ref{fig:dimerization}, we plot the expectation value of the staggered-flux operator $\hat{\tilde{T}}(\pi) = \frac{1}{N_s} \sum_i (-1)^i  T_{i_\ell} $, where we remind that $N_s$ is the number of sampled sites after removing the edges. Using MPS, we calculate the  order parameter ${\tilde{T}}(\pi)=\langle g|\hat{\tilde{T}}(\pi)|g\rangle$, which remains non-zero in a wide region of  parameter space  extending to larger values of $h/t$. 

We will now provide evidence that the BOW phase at $h\neq 0$ is still a SPT phase even when introducing quantum fluctuations in the gauge fields. In the presence of the electric field term, which generates  non-trivial long-range interactions between particles, we can no longer resort to topological band theory to define a bulk topological invariant from the single-particle spectrum~\eqref{eq:zak}. We shall instead characterize the topological features of the system by studying observables that are suitable to an interacting quantum many-body system. Indeed, we remark that while at $h=0$ the gauge-fields can be treated as a background for the fermions, now they are strongly intertwined. As a first consequence, we cannot rely on chiral symmetry anymore: at $h=0$, we could disregard the role of the magnetic term $J$ as a mere constant contribution, once the background flux configuration was fixed, and the chiral symmetry of the system was solely acting on the fermionic degrees of freedom. At $h\neq0$, chiral symmetry would need to correspond to some unitary operator $\hat{O}$ involving both matter and gauge degrees of freedom, such that $\{H,\hat{O}\} = 0$. By inspection of Eq.~\eqref{eq:totspin}, we conclude that such operator does not exist. Moreover, the aforementioned faster modulation associated with $\tilde{k}=\frac{\pi}{2}$ and resulting in a four-site periodicity generally leads to the breakdown of bond-inversion symmetry. However, we observe a pattern of gauge-fluxes $\langle T_{i_\ell}\rangle$ that alternates between a strong (S) and two weak ($\rm{W_1,W_2}$) links: $ \rm{SW_1SW_2}$ - see panel (a) of Fig.~\ref{fig:local_berry}, and preserves inversion around weak bonds $I_{\rm{W}}$ only, providing symmetry protection. As previously discussed in the case of $h=0$ for which, in a finite open chain, the topological ordering (WS) appears as an excitation above the trivial ordering (SW), this is also the case at $h\neq0$.

As a probe of the topological properties of the ground state, we examine the quantization of the local Berry phase introduced by Hatsugai~\cite{hatsugai2006quantized}, which serves as a topological invariant to characterize interacting many-body SPT phases and provides the usual definition of the Berry phase with a notion of locality. Specifically, one introduces a local perturbation on which the Hamiltonian depends periodically, which does not explicitly break the symmetry providing topological protection. For instance, a local phase-twist of the hopping amplitude across a weak bond $t_{i_\ell} \to t_{i_\ell} {\rm e}^{{\rm i \theta}}$ preserves inversion symmetry around it. The ground state of the system $\ket{\psi(\theta)}$ develops a Berry phase under a cyclic variation of $\theta$
\begin{equation}
    \label{eq:locberry}
   \gamma_{i_\ell}= -{\rm i} \int_\mathcal{C} \langle\psi(\theta) \vert \partial_\theta \vert \psi(\theta)\rangle
\end{equation}

\ni where $\mathcal{C}$  is the loop traversed by $\theta$ and the subscript $i$ denotes the position at which the perturbation is applied. In general, a phase transformation of the state $\vert \psi(\theta)\rangle \to e^{i\Omega(\theta)} \ket{\psi(\theta)}$ results in a change of ${\rm i} \, \gamma_\mathcal{C}$ by $\int_\mathcal{C} d\Omega$, evidencing the need for a gauge-fixing. This is achieved by considering a state $\ket{\phi}$ as a phase reference, provided that it is single-valued along the closed path $\mathcal{C}$ and that $\langle \psi(\theta) \vert \phi \rangle \neq 0$, and defining $\ket{\psi_\phi(\theta)} = \frac{\langle \psi \vert \phi \rangle}{\abs{\langle \psi \vert \phi \rangle}} \ket{\psi}$. The single-valuedness of the reference state $\ket{\phi}$ ensures that that the gauge choice is regular and that the Berry phase $\gamma_{i_\ell}^\phi$ associated to $\ket{\psi_\phi(\theta)}$ is defined up to an integer multiple of $2 \pi$. In the presence of an anti-unitary symmetry of the Hamiltonian, if the ground state is gapped and unique along the whole path $\mathcal{C}$, the Berry phase is quantized $\gamma_{i_\ell}^\phi = 0,\pi \, {\rm mod} \, 2 n\pi$ and can serve as a probe of topological order. Numerically, the parameter loop $\mathcal{C}$ is discretized in $N_{\mathcal{C}}$ points $\theta_n = \frac{2\pi n}{N_\mathcal{C}}$, $n=1,\dots,N_\mathcal{C}$ and the local Berry phase is approximated by 
\begin{equation}
    \gamma_{i_\ell}^\phi = {\rm Arg} \prod_{n=1}^{N_\mathcal{C}} \langle \psi^U_{n} \vert \psi^U_{n+1} \rangle, 
    \end{equation}
    with $\ket{\psi^U_n} = \ket{\psi_n} \langle \psi_n \vert \phi \rangle$, and $\ket{\psi_n}$ being the ground state of $H(\theta_n)$. Such expression is independent of the reference state $\ket{\phi}$ and recovers the Berry phase~\eqref{eq:locberry} as $N_\mathcal{C} \to \infty$. 

At $h=0$, paralleling Hatsugai's construction for a generalized SSH model~\cite{hatsugai2006quantized}, one can invoke particle-hole symmetry $\Theta = KU_\Theta$ to infer the quantization of $\gamma_{i_\ell}$, where $K$ is complex conjugation, while $U_\Theta$ is the unitary operator that maps $c_{2n} \to c_{2n}^\dagger$ and $c_{2n+1} \to -c_{2n+1}^\dagger$. It is straightforward to see that $[H(\theta),\Theta] = 0$. However, since $\Theta$ maps $n_i \to 1-n_i$, the Gauss'law operator $G_i = P^x_{i_\ell-1} (-1)^{n_i} P^x_{i_\ell}$ at site $i$ changes sign: particle-hole symmetry maps the choice $G_i = 1$ to $G_i = -1$. At $h=0$ this is unimportant, as all gauge sectors are equivalent and the sign change can be undone by local flips of $P^x$: for instance, this is achieved by the action of $U_z = \prod_{n \geq 0} {\rm exp}\{{\rm i} \pi S^z_{i_\ell - 1- 2n}\}$. When $h\neq0$, this symmetry is broken, as $U_z$ would restore $G_i = 1$ $\forall i$, but would also map $S^x_{i_\ell-1-2n} \to- S^x_{i_\ell-1-2n}$, $\forall n \geq 0$. For $h\neq0$, we can rely on the anti-unitary combination $K I_{\rm W}$, $[H(\theta),K I_{\rm W}] = 0$ to ensure that $\gamma_{i_\ell}$ is quantized on weak bonds.
\begin{figure}
    \centering
    \includegraphics[width=\linewidth]{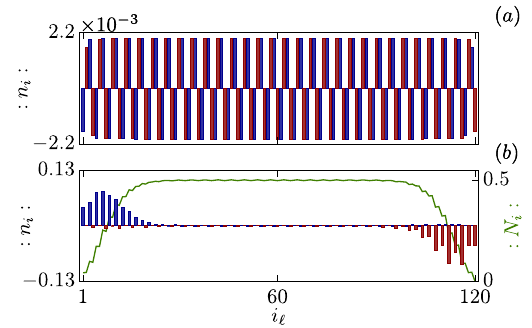}
    \caption{\textit{CDW and edge states}. The excess charge density $:n_i: = \langle n_i \rangle - \nu$ with respect to half-filling $\nu=\frac{1}{2}$ is evaluated in the trivial configuration realized in the ground state of the system (a) and in the topological state obtained after applying the pinning potential~\eqref{eq:pinning} $V(\lambda = 5 \cdot 10^{-3} \, t)$ (b). In both cases, we considered a chain of $L=120$ sites and $J/t = 0.2$, $h/t = 0.1$. In the first case, a very tiny CDW pattern manifests oscillations with the wave-vector $\tilde{k} = \frac{\pi}{2}$. In the topological configuration, charge-fractionalization occurs, manifest in the accumulation (depletion) of half a charge localized at the left (right) edges of the chain. This is evidenced by the solid green line, which refers to the cumulative excess charge $:N_i: = \sum_{j \leq i} :n_j:$ (second y-axis on the right hand side of panel (b)).}
    \label{fig:density}
\end{figure}

Nevertheless, we successively twist both weak and strong bonds along the chain, discretizing the loop in parameter space in $N_\mathcal{C} = 10$ points. The ground state at $\theta=0$ is chosen as our reference state $\ket{\phi}=\ket{g}$, and we make sure that the overlap $\langle \psi_n \vert g\rangle$ is non-vanishing at every $n \in \{1,\dots,N_{\mathcal{C}}\}$. We find that the local Berry phase is $0$ on all weak bonds at even numbered positions, and is very close to $\pi$ as the perturbation is applied to the odd numbered strong bonds ($\abs{\gamma_S/\pi} \sim 0.99$) - see Fig.~\ref{fig:local_berry}. The almost-perfect Berry phase quantization on strong bonds might suggest that there is another anti-unitary symmetry, rather than $KI_W$, supporting it and it will be interesting if future works can identify it.

The alternating pattern of the local Berry phase can only be reversed at a quantum phase transition as the energy gap closes and the roles of weak and strong bonds is exchanged. To demonstrate it, we add a tiny pinning potential of the form 
\begin{equation}
    \label{eq:pinning}
    V_{\rm pin}(\lambda) = -\lambda \sum_i \left[ \cos(\pi i) - \sin(\frac{\pi}{2} i) \right] \Delta_{i_{\ell}} \, ,
\end{equation}

\ni with $\lambda \ll t,h,J$, that selects one of the two possible bulk topological orderings ($\rm{W_1 S W_2 S}$) , resulting in an opposite alternation of $\gamma_{i_\ell}$ - see Fig.~\ref{fig:local_berry}. In the pinned topological state, consecutive bonds of the same kind appear as domain walls in the flux pattern $\langle T_{i_\ell} \rangle$, localized at the edges of the chain, and spread out due to the quantum fluctuations induced by $h$. At their positions, the local Berry phase becomes undefined, pointing at a gap closing as the parameter loop $\mathcal{C}$ is traversed, and fractional charges localize: by plotting the excess charge density with respect to half-filling $:n_i: = \langle n_i \rangle - \nu$, we notice an excess (depletion) of half a particle at the left (right) edge, which have support on different sublattices - see Fig.~\ref{fig:density}. A more thorough discussion of charge fractionalization at localized defects and of its implications will be the subject of the next section. We note here that explicitly breaking chiral symmetry by turning on a nonzero electric field strength $h$, lifts the topological protection of the edge states, which are no longer guaranteed to be at zero energy. In this case, the SPT order is protected by a crystalline symmetry, namely inversion with respect to weak bonds $I_{\rm W}$, which is essential to ensure the quantization of the local Berry phase thereby defined. An analogous situation is observed in $\mathbb{Z}_2$ Bose-Hubbard models~\cite{gonzalezcuadra2019symmetry,gonzalezcuadra2020zn}.

We remind that turning on $h\neq0$ produces a faster modulation of the gauge-fluxes $T_{i_\ell}$ and enlarges the unit cell to four sites. The onset of long-range interactions between particles, induced by the electric field term, also reflects in a very weak charge density wave (CDW) with the same period, associated to the wave-vector $\tilde{k}=\frac{\pi}{2}$ - see Fig.~\ref{fig:density}. This tetramerization, is an induced order, which vanishes outside of the dimerized region.

\begin{figure}
    \centering
    \includegraphics[width=\linewidth]{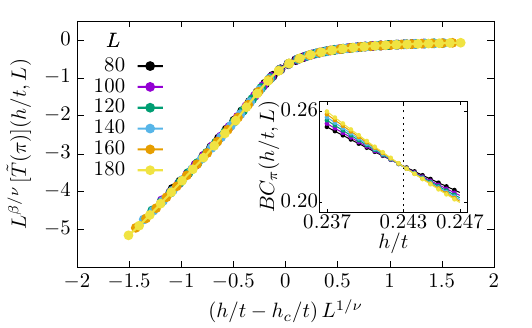}
    \caption{\textit{Finite-size scaling analysis.} A finite-size scaling analysis of the order parameter $\tilde{T}(\pi)$ and of the associated Binder cumulant $BC_\pi$ is performed at $J/t = 0.25$. The critical point $h_c$ and the correlation length critical exponent $\nu$ are found by fitting the Binder cumulant to a linearized scaling function (see text). A successive fit provides the critical exponent $\beta$ of the order parameter. In the main figure, we show the data collapse obtained for the order parameter with the fitted parameters $h_c \sim 0.243,\nu\sim2.464,\beta\sim 0.828$, for curves corresponding to system sizes raging from $L=80$ to $L=180$ sites, in jumps of $20$. In the inset, we show the crossing of the Binder cumulant curves at the critical point $h_c$ (marked by the dashed vertical line), for the same range of system sizes and in a more focused interval around the critical point (to ensure a better visualization, we do not show all the sampled points, but plot them every three).}
    \label{fig:scaling}
\end{figure}
To delimit the region of parameter space manifesting long-range TBOW, we have performed a finite-size scaling analysis to locate the critical points and estimate the critical exponents of the transition. In particular, we relied on the scaling properties of the Binder cumulant~\cite{binder1981finite,binder1984finite} associated to the order parameter $\tilde{T}(\pi)$, defined as 
\begin{equation}
    BC_\pi = 1-\frac{\langle g|\hat{\tilde{T}}(\pi)^4 |g\rangle}{3 \langle g | \hat{\tilde{T}}(\pi)^2 |g\rangle^2},
    \end{equation}
    where the hat on $\hat{\tilde{T}}(\pi)$ distinguishes the operator from the order parameter $\tilde{T}(\pi)$. The Binder cumulant is a dimensionless quantity that, in the scaling regime, behaves as 
    \begin{equation}
    BC_\pi(g,L) = F_{BC}(L^{1/\nu}(g-g_c)),
    \end{equation}
    such  that curves in function of a microscopic coupling  $g$ at different system sizes $L$ cross at the critical point $g_c$. 
    Considering large enough system sizes $L$ and a dense grid of points around the location of the observed pairwise crossings between the cumulant curves, we fit the position of the critical point $g_c$ and the exponent $\nu$ to a linearized scaling function $F_{BC} \approx A+ B \, (g-g_c) L^{\frac{1}{\nu}}$ for large $L$ and small $(g-g_c)$. The extracted parameters allow us to also estimate the critical exponent $\beta$ characterizing the scaling of the order parameter $[\tilde{T}(\pi)](g,L) = L^{-\beta/\nu} F_{\tilde{T}(\pi)}(L^{\frac{1}{\nu}}(g-g_c))$ and providing data collapse - see Fig.~\ref{fig:scaling} for the scaling analysis at $J/t = 0.25$. Second-order quantum phase transitions separate the BOW phase from an homogeneous one in which $\langle T_{i_\ell} \rangle$ is ferromagnetically ordered and translational invariance is restored. The extracted critical points are also shown in Fig.~\ref{fig:dimerization} and precisely delimit the boundaries of the inhomogeneous TBOW region. We also mention here that the correlation length critical exponent $\nu$ is not constant, but rather changes along the critical line. As we approach smaller values of $h/t$ towards the left and right edges of the symmetry-broken area of parameter space, we fail to obtain clean Binder cumulant's crossings. At the same time, while the transition becomes of first order at $h=0$ and $J=J_{c,1},J_{c,2}$ - see Eq.~\eqref{eq:fo_crit}, no signature of discontinuous behaviour is found. Further analysis would thus be needed to determine the nature of the transition.

    We have thus delimited a finite region of the parameter space in which SSB and SPT orders coexist. In spite of being different kinds of order, they are not exclusive, as long as the spontaneously broken symmetry does not coincide with the one providing symmetry protection. We also note that, besides resorting to local order parameters, some SPT phases can be characterized by string-order parameters, for instance in dimerized SSH-type models~\cite{Bahovadinov_2019}. While this would be the case in the $h=0$ limit of the studied theory, understanding whether this remains true at $h\neq0$ would require further investigations.

\end{subsection}

\section{\bf Deconfinement of fractionally-charged solitons}
\label{sec:solitons}
In the previous sections, we have studied the phase diagram of our multi-link $\mathbb{Z}_2$ gauge theory in the maximal spin sector for each link~\eqref{eq:totspin}, exploring the interplay between the physics of SSB and the symmetry-protected topological phases induced by the breakdown of translational invariance to its subgroup of four-site translations $\mathbb{Z}_4$. In this section, we will investigate the effects of doping the system above half-filling, linking topology and SSB to the physics of confinement.

In our discussion of the $h=0$ limit, we could clearly draw an analogy between our gauge theory and the physics of Peierls insulators described by SSH-like models. In these systems, inhomogeneous lattice deformations are considered to be a fixed background for the fermions to move. In the presence of domain walls that interpolate between the  SSB  orders, particles can remain bound  and acquire a fractional charge~\cite{PhysRevD.13.3398}. This phenomenon of charge-fractionalization stems from symmetry protection and is a realization of the bulk boundary correspondence~\cite{bernevig2013topological,asboth2016lecture}, linking the value of the fractional charge to the difference in the bulk topological invariants of the distinct sublattices. 

Static defects, however, do not capture the interplay between SSB and topological protection, but rather force an already symmetry-broken configuration. Modeling dynamical lattice deformations, as has been achieved for instance in the aforementioned $\mathbb{Z}_2$ Bose-Hubbard models~\cite{gonzalezcuadra2019symmetry}, is essential to clarify the nature of these topological defects and their interplay with dynamical matter. In particular, it was shown that domain walls can spontaneously form in the ground state of the system when doped above commensurate fillings and spread out, due to quantum fluctuations, to form topological solitons.
\begin{figure}
    \centering
    \includegraphics[width=\linewidth]{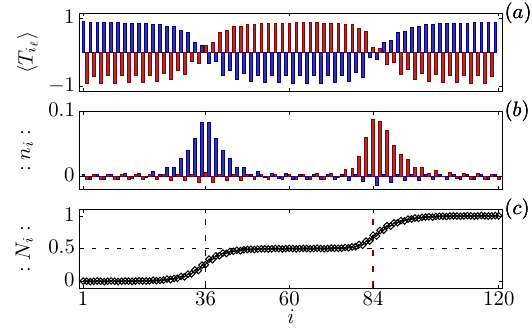}
    \caption{\textit{Topological fractionally-charged solitons.} A soliton/anti-soliton pair spontaneously arises in ground state of the system upon doping it with one particle and their positions are pinned through the addition of the pinning potential $V^{i_1,i_2}_{\rm pin}(\epsilon=0.05 \, t)$ of Eq.~\eqref{eq;soliton_pin}, with pinning centers $i_1 = 36$, $i_2 = 84$. In panel (a) we plot the gauge flux distribution $T_{i_\ell}$, evidencing how the two solitons interpolate between trivial and topological bulk orderings. In panel (b), the profile of the excess charge density $:n_i: = \langle n_i \rangle - 1/2$ shows that the charge excess is localized around the pinned soliton centers. Panel (c) displays the cumulative excess charge $:N_i: = \sum_{j \leq i} :n_j:$, evidencing how each soliton binds exactly half a fermion. Blue and red dashed vertical lines highltight the position of the pinning centers, while the horizontal dashed black line is added to demark the value $:N_i: = 0.5$. The Hamiltonian parameters are chosen to $J/t = 0.3,h/t=0.1$.}
    \label{fig:solitons}
\end{figure}
We will now demonstrate that the formation of topological solitons is also a feature of our multi-link gauge theory and we will discuss its fundamental implications on the physics of confinement. Remarkably, we find that fractional charges bound at these topological defects emerge as deconfined quasi-particles, which can be separated at any arbitrary distance without being subject to a confining potential, even in the presence of a non-zero electric field strength $h$ mediating long-range interactions. A similar mechanism of soliton-induced deconfinement was demonstrated for a $\mathbb{Z}_2$ LGT defined on a Creutz-ladder~\cite{gonzalez2020robust}, although with integer charges. To the best of our knowledge, this is the first time that the deconfinement of fractional charges bound to topological defects is observed in a quasi-one-dimensional $\mathbb{Z}_2$ LGT.

To address this phenomenon, we consider a half-filled chain of $L=120$ sites, choosing the Hamiltonian parameters within the TBOW region, and we dope it with one extra particle. According to the Gauss' law constraint, this is not possible unless an odd number of background charges is introduced into the system. Hence, we deform the Gauss' law on the left edge to account for a static background charge at site $1$: $G_1 \to -G_1$. Upon doping the system, a soliton/anti-soliton pair is spontaneously created, interpolating between trivial and topological bond orderings. While at $h=0$ two sharply localized domain walls would form, at $h\neq0$ the electric field terms induces quantum fluctuations of the gauge flux $\langle T_{i_\ell} \rangle$, causing the domain walls to delocalize into solitons. Soliton and anti-soliton have support on different sublattices and each of them binds quasiparticles with fractional charge $q_f = \frac{1}{2}$ - see Fig.~\ref{fig:solitons} . With the aim of controlling their position along the chain, following a similar strategy to that presented in~\cite{gonzalezcuadra2019symmetry}, we have introduced a pinning potential, which is more conveniently expressed in terms of the two sets of Pauli operators $\boldsymbol{\rho}_{i_\ell}$, $\boldsymbol{\tau}_{i_\ell}$ which we had previously introduced to rewrite spin-$\frac{3}{2}$ operators in the common eigenbasis of $P^x_{i_\ell}$ and $(S^z_{i_\ell})^2$. Namely:
\begin{equation}
    \label{eq;soliton_pin}
    V^{i_1,i_2}_{\rm pin}(\epsilon) = \epsilon \sum_{j \in \{i_1,i_2\}} [\tau^x \otimes \mathbb{I}_2 ]_{j_\ell-1} [\tau^x \otimes \mathbb{I}_2 ]_{j_\ell} \, ,
\end{equation}

\ni where $i_1,i_2$ are the pinning centers and $\epsilon$ is small enough not to alter the energy spectrum, but big enough to localize the solitons at the chosen positions. 
\begin{figure}
    \centering
    \includegraphics[width=\linewidth]{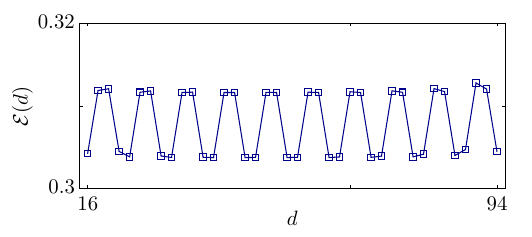}
    \caption{\textit{Deconfinement of fractional  solitons}. The soliton and anti-soliton configurations, that arise when an extra fermion is introduced into the system above half-filling, are pinned at symmetrical positions $i_1$ and $i_2 = L-i_1$ about the central site of a chain of size $L=120$. As the distance $d=i_2-i_1$ is gradually increased, we find the ground state energy for this system of $N_p+1$ particles $E_g^{N_p+1}(d)$, where $N_p = L/2$ and plot the difference from $E_g^{N_p}$, defined as $\mathcal{E}(d) = E_g^{N_p+1}(d) - E_g^{N_p}$. We start from $d=16$, to guarantee well separated solitons. The resulting energy landscape manifests the deconfinement of the fractional quasi-particles bound at the defects, remaining almost constant across all distances. An oscillating behaviour with a four-site periodicity reflects the SSB of translational invariance to the group $\mathbb{Z}_4$ and are an example of Peierls-Nabarro barriers. Large distances $d$ are cut, as the pinned solitons overlap with the boundaries of the chain and are repelled.}
    \label{fig:deconfinement}
\end{figure}
We take advantage of this pinning protocol to investigate the spatial dependence of the interaction between the solitons. To this aim, we localize each of them symmetrically with respect to the site $L/2$, far enough from the center to ensure that they are well separated, and we calculate the ground state energy as we vary their distance. The resulting energy landscape, shifted by the ground state energy at exactly half filling, is reported in Fig.~\ref{fig:deconfinement} and demonstrates an oscillating behaviour as the pinning centers are moved within the four-site unit cells, compatibly with the residual $\mathbb{Z}_4$ symmetry after SSB. These oscillations are an example of Peierls-Nabarro barriers, i.e. energy barriers determined by the breakdown of lattice translational invariance~\cite{peierls1940the,nabarro1947dislocations}. Remarkably, even in the presence of a string tension term inducing long-range interactions between fermions, quasi-particles carrying fractional charge and bound to the soliton/anti-soliton pair are deconfined, as they can be pulled apart to arbitrary distances without incurring in a linearly growing energy cost or any string breaking phenomenon. Outside of the topological BOW region of the phase diagram, confinement is restored: adding an extra particle above half-filling leads to a density accumulation at the left edge, growing with the confining strength $h/t$, and reflecting a confined bound-state with the added background charge - see Fig.~\ref{fig:n1} for a plot of $\langle n_1\rangle$ as a function of $h/t$. 
\begin{figure}
    \centering
    \includegraphics[width=\linewidth]{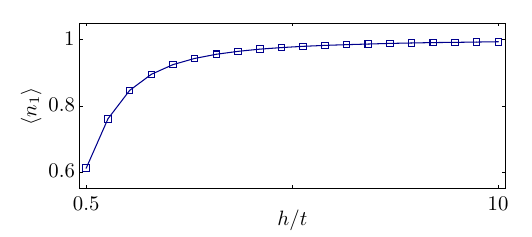}
    \caption{\textit{Background charge confinement}. Outisde of the symmetry-broken TBOW region, the doping particle leads to a density accumulation at the location of the background charge $\langle n_1 \rangle$, pointing at the formation of a confined bound state. We plot $\langle n_1 \rangle$ as a function of $h/t$ at $J/t = 0.2$.}
    \label{fig:n1}
\end{figure}
\end{section}

\begin{section}{\bf Conclusions and outlook}
In this paper we have introduced a multi-link $\mathbb{Z}_2$ LGT supporting minimal elementary plaquettes, which are loops enclosed by pairs of links - see Fig.~\ref{fig:z2_scheme}. The configurations attained by the $\mathbb{Z}_2$-valued gauge magnetic flux threading each loop, intertwining with the particle's gauge-invariant tunneling dynamics, have been shown to be the driving mechanism leading to a Peierls-like transition, when the number of links per bond is odd. As opposed to the case $N_b$ even~\cite{domanti2025dynamical}, for which completely destructive interference, resulting from dynamical $\pi$ flux configurations, cage particles into disconnected portions of the lattice, when $N_b$ is odd not all pairs of interfering paths can enclose a $\pi$ flux. This results into a state-dependent tunneling amplitude for the fermions, which is enhanced(diminished) the more(less) $0$-flux states are present, supporting constructing interference. Notably, this mechanism leading to Peierls instability is perfectly compatible with the local $\mathbb{Z}_2$ gauge symmetry, when discussed in terms of the gauge-invarant Wilson loops $W_{i_\ell,\circlearrowleft}$ quantifying the gauge-flux configurations. A naive mean-field description is instead precluded by gauge symmetry, as the mean-field tunneling amplitudes $t \langle S^z_{i_\ell} \rangle$ would vanish identically to fulfill gauge invariance~\cite{elitzur1975impossibility} - see Eq.\eqref{eq:totspin}.

At zero electric field $h=0$, we have demonstrated the existence of inhomogeneous phases exhibiting a periodic spatial ordering of the gauge-fluxes, with a wave-vector equal to the Fermi momentum, which survive in the presence of the quantum fluctuations driven by a small electric field $h>0$ - see Sec.~\ref{sec:peierls} and Fig.~\ref{fig:cp_analysis} therein. Such inhomogeneous phases are stabilized in incompressible regions of parameter space at the fermion fillings $\nu=\frac{1}{2},\frac{2}{3}$. Here, the system manifests SSB of the lattice translational symmetry and develops a long-range ordered pattern of the gauge-invariant bond kinetic operators $\langle B_{i_\ell} \rangle$, i.e. a bond-order wave (BOW). At half-filling, the breakdown of translational symmetry is manifest as a spontaneous dimerization of the gauge-flux and bond strengths which, at $h=0$, is described by a SSH-like model~\cite{su1979solitons}. 

For $h>0$ gauge-fields are no more a mere static background but the electric field provides particles and gauge fields with a non-trivial dynamics. Its effects are two-fold: on the one hand, quantum fluctuations of the Wilson loops are set that result in a four-site-periodic sub-modulation of the observables inside the symmetry-broken region; on the other, a flux-dependent string tension term is induced that supports long-distance particle interactions. The long-range ordered BOW phase is shown to coexist with symmetry-protected topological phenomena, which  manifests in the quantization of the local Berry phase and in the appearance of localized edge states in the topological ground-state - see Sec.~\ref{sec:tbow}. Such SPT phase is separated by second order phase transitions from gapless homogeneous phases in which the lattice translational invariance is restored. There, a strong electric field results in the confinement of fermions into tightly-packed dimers, which move freely  as neutral composite excitations~\cite{kebric2021confinement,domanti2025dynamical}. 

As the phase boundaries are crossed into the topological BOW region, a mechanism of  deconfinement emerges from charge fractionalization. A background charge is located to one edge of the chain deforming the Gauss' law and allowing for an extra particle to be added to the system above half-filling - see Sec.~\ref{sec:solitons}. Upon doping, pairs of domain-walls emerge that interpolate between the symmetry-broken trivial and topological orderings of the gauge-flux and delocalize due to the electric-field-induced quantum fluctuations into soliton and anti-soliton pairs. Two quasiparticles with fractional charge $q_f=\frac{1}{2}$ appear that are bound at the soliton centers and are deconfined. Pinning the solitons' positions at increasing distances only costs the energy necessary to overcome Peierls-Nabarro barriers~\cite{peierls1940the,nabarro1947dislocations}, arising from the breakdown of translational invariance. No confining force is present that binds the fractional soliton/anti-soliton pairs which thus are deconfined and can be arbitrarily spaced.

The phenomena described in this manuscript rely on the dynamical Aharonov-Bohm effect induced by gauge-invariant configurations of the magnetic flux through the elementary loops enclosed by pairs of links in the multigraph geometry. We thus expect similar, but also richer phenomena to occur as more complex gauge groups are considered, for instance $\mathbb{Z}_N$, allowing for gauge-invariant, $\mathbb{Z}_N$-valued fluxes.

The structural simplicity of the multi-link $\mathbb{Z}_2$ gauge theories studied in our work, where Wilson loops are weight-2 operators and magnetic terms reduce to two-body Ising interactions, makes this type of models inetresting for implementation in analog quantum simulators, such as trapped ions, as they avoid higher-weight plaquettes requiring four-body interactions. The two-body Ising interactions can be realized in trapped ions according to various schemes through state-dependent forces, and have already been demonstrated with high precision in seminal experiments~\cite{RevModPhys.93.025001}. Together with the recent realization of multi-link gauge structures in this platform~\cite{saner2025real}, our results indicate that the interplay between dynamical gauge fluxes, Peierls instabilities, and fractionalized solitonic excitations might be explored experimentally in the future. The results presented in this work highlight how these experiemts could test the non-trivial interplay between global symmetry breaking and exact local gauge invariance, the emergence of symmetry-protected topological phases, and the deconfinement of fractionalized charges within a unified lattice gauge framework.

\end{section}

\acknowledgements We acknowledge support from the European
Unions Horizon Europe research and innovation programme
under grant agreement No 101114305 (MILLENION-
SGA1 EU Project), from PID2021-127726NB-I00 and
PID2024-161474NB-I00 (MCIU/AEI/FEDER,UE), and from
QUITEMAD-CM TEC-2024/COM-84. We also acknowledge support from the Grant IFT Centro de Excelencia Severo Ochoa CEX2020-001007-S funded by
MCIN/AEI/10.13039/501100011033, and from the CSIC Research Platform on Quantum Technologies PTI-001. 

\bibliographystyle{apsrev4-1}
\bibliography{bibliography}

\end{document}